\documentclass[fleqn,usenatbib]{mnras}

\usepackage{newtxtext,newtxmath}

\usepackage[T1]{fontenc}

\DeclareRobustCommand{\VAN}[3]{#2}
\let\VANthebibliography\thebibliography
\def\thebibliography{\DeclareRobustCommand{\VAN}[3]{##3}\VANthebibliography}

\usepackage{graphicx}	
\usepackage{amsmath}	
\usepackage{graphicx}	
\usepackage{amsmath}	

\usepackage{xcolor}
\usepackage{hyperref}
\usepackage{url}
\usepackage{verbatim}
\usepackage{booktabs}
\usepackage{tabularx}
\usepackage{caption}

\usepackage{xspace}
\usepackage[normalem]{ulem}

\newcommand{\cms}{\ensuremath{\rm cm\,s^{-1}}}
\newcommand{\kms}{\ensuremath{\rm km\,s^{-1}}}
\newcommand{\ms}{\ensuremath{\rm m\,s^{-1}}}
\newcommand{\Bl}{$B_{\rm l}$\xspace}

\title[Longitudinal Magnetic Field in RV Surveys]{The Mean Longitudinal Magnetic Field and its Uses in Radial-Velocity Surveys}

\author[F. Rescigno et al.]{F. Rescigno$^{1}$\thanks{E-mail: f.rescigno@bham.ac.uk},
A. Mortier$^{2}$,
X. Dumusque$^{3}$,
B. S. Lakeland$^{1}$,
R. Haywood$^{1}$,
N. Piskunov$^{4}$,
B. A. Nicholson$^{5}$,
\newauthor
M. L\'opez-Morales$^{6}$,
S. Dalal$^{1}$,
M. Cretignier$^{7}$,
B. Klein$^{7}$,
A. Collier Cameron$^{8,9}$,
A. Ghedina$^{10}$,
M. Gonzalez$^{10}$,
\newauthor
R. Cosentino$^{10}$,
A. Sozzetti$^{11}$,
S. H. Saar$^{6}$
\\
$^{1}$Department of Astrophysics, University of Exeter, Stocker Rd, Exeter, EX4 4QL, UK\\
$^{2}$Department of Astrophysics, University of Birmingham, Edgbaston, Birmingham, B15 2TT, UK\\
$^{3}$Observatoire de Gene\'eve, Universit\'e de Gene\'eve, Chemin de Pegasi, 51, CH-1290 Versoix, Switzerland\\
$^{4}$Department of Physics and Astronomy, Uppsala University, Box 516, 75120 Uppsala, Sweden\\
$^{5}$Centre for Astrophysics, University of Southern Queensland, West St, Toowoomba, QLD, 4350, Australia\\
$^{6}$Center for Astrophysics | Harvard \& Smithsonian, 60 Garden Street, Cambridge, MA 02138, USA\\
$^{7}$Sub-department of Astrophysics, Department of Physics, University of Oxford, Oxford, OX1 3RH, UK\\
$^{8}$SUPA, School of Physics \& Astronomy, University of St Andrews, North Haugh, St Andrews, KY169SS, UK\\
$^{9}$Centre for Exoplanet Science, University of St Andrews, North Haugh, St Andrews, KY169SS, UK\\
$^{10}$Fundaci\'on Galileo Galilei - INAF (Telescopio Nazionale Galileo), Rambla J. A. F. Perez 7, E-38712 Bre\~na Baja (La Palma), Canary Islands, Spain\\
$^{11}$INAF - Osservatorio Astrofisico di Torino, Strada Osservatorio, 20 I-10025 Pino Torinese (TO), Italy\\
}

\date{Accepted XXX. Received YYY; in original form ZZZ}

\pubyear{2024}

\begin{document}
\label{firstpage}
\pagerange{\pageref{firstpage}--\pageref{lastpage}}
\maketitle

\begin{abstract}
This work focuses on the analysis of the mean longitudinal magnetic field as a stellar activity tracer in the context of small exoplanet detection and characterisation in radial-velocity (RV) surveys. 
We use SDO/HMI filtergrams to derive Sun-as-a-star magnetic field measurements, and show that the mean longitudinal magnetic field is an excellent rotational period detector and a useful tracer of the solar magnetic cycle. To put these results into context, we compare the mean longitudinal magnetic field to three common activity proxies derived from HARPS-N Sun-as-a-star data: the full-width at half-maximum, the bisector span and the S-index. The mean longitudinal magnetic field does not correlate with the RVs and therefore cannot be used as a one-to-one proxy. However, with high cadence and a long baseline, the mean longitudinal magnetic field outperforms all other considered proxies as a solar rotational period detector, and can be used to inform our understanding of the physical processes happening on the surface of the Sun.
We also test the mean longitudinal magnetic field as a "stellar proxy" on a reduced solar dataset to simulate stellar-like observational sampling. With a Gaussian Process regression analysis, we confirm that the solar mean longitudinal magnetic field is the most effective of the considered indicators, and is the most efficient rotational period indicator over different levels of stellar activity. This work highlights the need for polarimetric time series observations of stars.
\end{abstract}




\section{Introduction}

In the last 20 years, the radial-velocity (RV) method has been used to successfully detect and characterise hundreds of exoplanets, from blazing hot giants to rocky super-Earths. With the aim of finding potential Earth analogues that future missions such as the Habitable Worlds Observatory\footnote{Based on the studies for the Large UV/Optical/IR Surveyor \citep[LUVOIR:][]{LUVOIR2019}, and the Habitable Exoplanet Observatory \citep[HabEx:][]{Gaudi2020}} \citep[HWO:][]{Harada2024} or the Large Interferometer for Exoplanets \citep[LIFE:][]{Quanz2022} can observe in search of biosignatures, the community is now more than ever targeting rocky exoplanets in their stellar habitable zone.
The RV signals imprinted by these planets on the light of their host stars are however of the order of tens of \cms \,and until recently have been out of reach for most precise spectrographs.

As instrumentation reaches the required precision, the biggest challenge we now face for the detection of Earth analogues and the accurate measurements of their masses is stellar variability \citep{Saar1997,Lindegren2003, Meunier2010, Dumusque2011,Fischer2016,Crass2021,Meunier2021}. The RV variations generated by activity on the surface of host stars are often of the order of several \ms and they can easily mimic or completely drown planetary oscillations. Stellar-induced signals, in fact, often dominate the RVs of Sun-like stars. These signals are still challenging to model as they affect the time series over multiple timescales, from minutes to years. Moreover, the longer baselines required for disentangling Earth-like signals introduce a further source of "noise": magnetic cycles.
Over the years stars are expected to undergo similar activity cycles \citep[e.g.][]{Olah2009} to what the Sun experiences, with years of maxima, where activity is much stronger and more significantly modulated by stellar rotation, and stretches of minima, where activity-induced variations are much weaker and non-rotationally-modulated effects dominate (e.g., granulation and supergranulation). Understanding and modelling these long-term cycles is often necessary for a comprehensive characterisation of planetary systems, in particular in the case of possible wide companions. A contemporaneous effort towards the confirmation of outer planets may in fact be vital for the detection of Earth analogues. In fact, recent studies have shown that the formation of inner Earths is dependent on the presence of quickly-accreted long-orbit gas giants \citep{Morbidelli2022}.

Great care is then required when accounting for and modelling stellar activity in order to obtain accurate orbital solutions and to precisely determine planetary masses. 
A common approach is to use Gaussian Processes (GP). GP regression has proven to be an effective tool for modelling stellar activity \citep[e.g.,][]{Haywood2014, Rajpaul2015, Faria2016, Serrano2018, Barros2020}. It however has its limitations. In particular, its ability to predict stellar variability is strongly reliant on an accurate detection of the stellar rotational period \citep{Nicholson2022}. A precise determination of the periodicity of the stellar-induced RVs is vital to correctly differentiate them from Keplerian signals \citep{Bortle2021}, and to compute accurate masses \citep{Blunt2023,Dalal2024}. In fact, confirming the presence of non-transiting planets can be particularly challenging when the stellar rotational period is similar to the orbital period of the planet candidate \citep{Nava2022}. An inaccurate rotational period can also have significant direct impacts on the derived best results for all the other kernel hyperparameters that are less reliably tied to physical processes and are therefore much harder to interpret correctly.

The accurate detection of the stellar rotation period is vital for many other areas of astrophysics beside exoplanet characterisation. As an example, \cite{Irving2023} studied the relationship between stellar rotation periods and magnetic cycle amplitudes, as well as between the ratio of rotation and cycle periods and the stellar Rossby number. \cite{McQuillan2014} highlighted a bimodality in the rotation period-temperature relation of more than 30,000 \textit{Kepler} targets. \cite{Amard2020} studied the contribution of stellar metallicity to the decay of rotational periods with age. On the whole, accurate and precise measurements of stellar rotational periods are at the basis of multiple current fields of study.

The stellar rotation period is often challenging to extract only from RVs \citep{Nava2020} or photometry, especially in times of low activity \citep{Aigrain2015}. We therefore use activity proxies as extra suppliers of information. Useful indicators are not sensitive to the Doppler shifts caused by the presence of planets. They instead map the spectral line distortions generated by the activity on the surface of the star or are generated by chromospheric emissions. Most commonly-used activity proxies are either extracted from the same spectra as the RVs, such as the S-index, or are derived from the shape of their cross-correlation function (CCF), such as the full-width at half-maximum (FWHM) and the bisector span (BIS). Nevertheless, even an analysis of these common activity indicators often fails to consistently measure the stellar rotation period \citep[e.g.,][]{Nava2022}. Therefore for the analysis of RVs and in particular for the detection of the stellar rotational period, especially over all stages of a star's magnetic cycle, a different tracer of activity is required.

\begin{figure*}
    \centering
    \includegraphics[width=15cm]{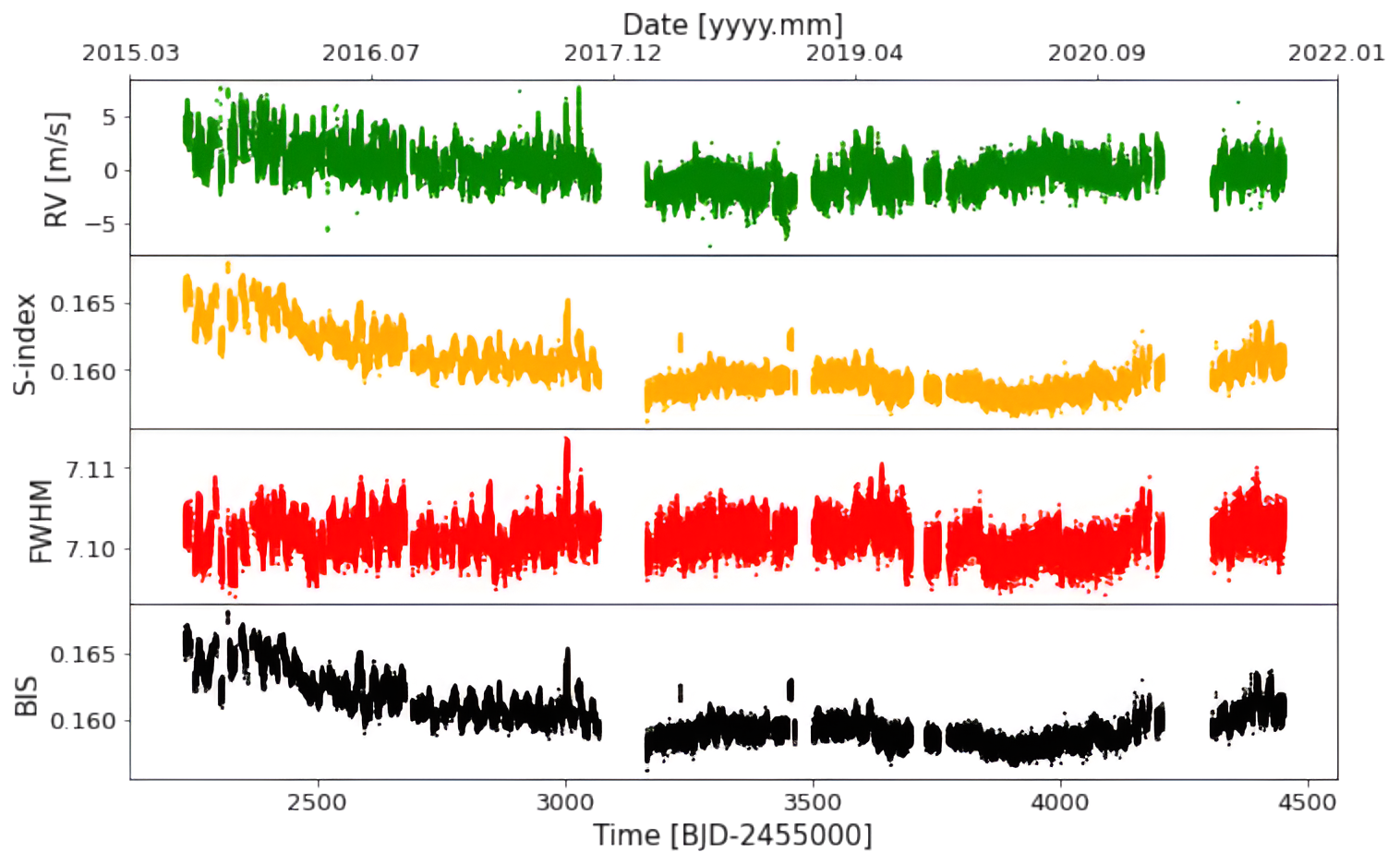}
    \caption{HARPS-N Solar telescope data. From the top, the corrected radial velocities in green, the S-index in orange, the full-width at half-maximum in red, and the bisector span in black. Uncertainties are included but too small to be visible.}
    \label{fig:HN}
\end{figure*}
\subsection{The Mean Longitudinal Magnetic Field}
\label{sec:bl}

Quantities that directly measure stellar magnetic field properties may be more useful for measuring stellar rotation and activity-induced RV variability.
\cite{Haywood2022} have shown that the unsigned (absolute) magnetic flux maps the stellar-induced RV variations better than any other activity indicator to date. However, measuring the absolute magnetic flux in stars that are not the Sun is extremely challenging \citep{Reiners2012}. Stellar magnetism is usually investigated with polarimetric observations (see \cite{Trippe2014} and similar reviews).

In the model of light as an electromagnetic (EM) wave, light polarisation is characterised by the four Stokes parameters: total intensity, $I$, two linear polarisation parameters, $Q$ and $U$, and one circular polarisation parameter, $V$. Stokes $Q$ corresponds to an EM wave propagating vertically or horizontally with respect to the line-of-sight. Stokes $U$ is an EM wave propagating on a plane on a $\pm 45^{\circ}$ angle from the $Q$ axis. Stokes $V$ describes light where there is a phase offset between the electric and magnetic parts of the EM wave, producing either right hand (clockwise) or left hand (anti-clockwise).
Typically, a single observation yields two Stokes parameters: $I$ and one of polarisations.
Light produced in the presence of a magnetic field becomes polarised through the Zeeman effect. In the case of measuring the polarisation of light produced by stars, $V$ traces the line-of-sight (longitudinal) component of the large-scale magnetic field, and $Q$ and $U$ measure the transverse components. While all components are needed to fully describe the magnetic field state, the measurement of linear polarisation ($Q$ and $U$) are more challenging for stars than circular ($V$). The observation of a single polarisation state ($Q$, $U$ or $V$) requires four sub-exposures, meaning that linear polarisation takes twice the telescope time to complete. Additionally, linear polarisation is a second order effect in the wavelength domain and, especially in the case of weak-fields, is a smaller component of the total intensity compared to $V$ \citep{Bagnulo2015}, and is more prone to spurious signals from surface reflections.
While progress has been made for the estimation of magnetic fields (and in particular of unsigned magnetic fluxes) for other stars \citep[see e.g., ][]{Lienhard2023,Kochukhov2023}, this remains challenging.

Given these difficulties, Stokes $V$ observations are currently the more efficient way study changes in large-scale stellar magnetic fields by measuring the net line-of-sight signed magnetic flux, also called longitudinal magnetic flux.
The mean longitudinal magnetic field, \Bl (sometimes also referred to as <$B_{\rm z}$>), is the line-of-sight projected component of the magnetic field vector averaged over the visible hemisphere of the star.
\Bl is related to the circular polarisation as \citep[e.g.][]{Landstreet1982}:
\begin{equation}
    \frac{V}{I} = g_{\rm eff} C_{\rm z} \lambda_{0}^{2} \frac{1}{I} \frac{dI}{d\lambda} B_{\rm l},
\end{equation}
where $g_{\rm eff}$ is the effective Land\'e factor, $I$ the intensity at wavelength $\lambda$, $\lambda_0$ is the average wavelength, and $C_{\rm z} = 4.67 \times 10^{-13}$ \AA$^{-1}$ G$^{-1}$.
The mean longitudinal field can then be expressed as the first order moment of the Stokes $V$ parameter as \citep{Donati1997}:
\begin{equation}
    B_{\rm l} = -2.14 \times 10^{11} \frac{\int \nu V(\nu) d\nu}{\lambda_{\rm av} g_{\rm av} c \int[I_{\rm c} - I(\nu)] d\nu},
\end{equation}
where $\lambda_{\rm av}$ and $g_{\rm av}$ refer to the average wavelength and the average Land\'e factor of the lines used to compute \Bl, and $I_{\rm c}$ is the continuum intensity. The integration limits over frequency $\nu$ are somewhat arbitrary and can change between analyses. They are selected wide enough to include all the information of the Stokes profiles but narrow enough to reduce the contribution of noise.

Analyses of \Bl time series have successfully determined rotation periods using traditional methods such as Lomb-Scargle periodograms.  Studies done with \Bl data from the near-infrared SPectropolarimètre InfraROUge (SPIRou: \citealt{Donati2020}) have detected rotational periods of chemically peculiar stars \citep{Babcock1949}, and M-dwarfs \citep[e.g.,][]{Landstreet1992, Donati2009, Klein2021, Fouque2023}. With these successes other polarimeters were also turned to similar analyses \citep[e.g.,][]{Hebrard2016, Nicholson2021, Marsden2023}.\cite{Yu2019} and later \cite{Donati2023} also introduced Gaussian Process regression to the modelling of \Bl in M-dwarfs.

For Sun-like stars, and most of the stars selected in RV surveys for exoplanet detection, the low observed projected rotational velocity \mbox{($v\sin{i}$ < 2 \kms)} makes the detection of magnetic fields difficult due to magnetic flux cancellation between opposite polarities.
To test these limits, \cite{Petit2008} observed a small sample of active Sun-like stars, and successfully detected their magnetic field. On a larger scale, the BCool magnetic survey \citep{Marsden2014} analysed spectropolarimetric data of 170 solar-type stars (F-, G- and K-type or FGK) collected between 2006 and 2013. They were able to detect the magnetic field in 1/3 of the sample, and of these stars 21 were Sun-like. With mostly a single observation per star, the survey reached precisions in \Bl of 0.2 G, and demonstrated that \Bl in quieter Sun-like stars is measurable with reasonable uncertainties. However, this survey and the majority of previous polarimetric surveys focused on obtaining mostly single snapshot observations. The great majority of FGK stars lack the time series of polarimetric data necessary to do period detection analysis. We therefore lack the data to analyse the longitudinal magnetic field as a stellar activity tracer. This work will present proof to the need of polarimetric time series stellar observation. In order to circumvent this limitation and as previous studies have done to better understand stellar variability and its dependence to other measurable quantities \citep{Haywood2016, CollierCameron2019, Haywood2022}, we turned to the best observed FGK star: the Sun.

The origin of \Bl, in solar science often also called Solar Mean Magnetic Field (SMMF) or General Magnetic Field (GMF), is still strongly debated. Some attribute the largest contribution to the signal of \Bl to the weak, large-scale magnetic flux over the entire visible disc (i.e., the background flux, or the "quiet" Sun flux) (e.g., \citealt{Severny1971, Xiang2016}). In fact, \cite{Bose2018} claim that 80\% of the signal of \Bl is generated by the background magnetic flux. Using resolved full-disc solar data, they partition the solar surface between sunspots, faculae and background. They then calculate the percentage variation of \Bl due to each region independently using the coefficient of determination method based on linear regression analysis. They found that there exists a clear correlation between \Bl and its component from only the background. They also found no correlation between the mean longitudinal magnetic field and the active regions filling factors. They concluded that the presence of active regions does not directly influence the structure of the signal in \Bl, but their location on the disc can influence the amplitude of the signal (as we will further explore in Section \ref{sec:full}).
These results are in opposition to the ones of others. For example, \cite{Scherrer1972} show that the largest correlation between \Bl and the interplanetary magnetic fields is reached when considering only the innermost fourth of the solar disc, which is more sensitive to active latitudes and therefore to active regions. Furthermore, \cite{Kutsenko2017} used a similar technique to \cite{Bose2018} on similar data, but recovered different results. They considered a magnitude threshold of 30 Mx cm$^{-2}$ (equal to 30 G) and found that the \Bl component derived from active regions contributed from 65 to 95\% of the total field. They therefore claim that \Bl is directly generated by magnetic flux concentrations, meaning spots, faculae and network. They assert that the strong rotational modulation measured in \Bl is a clear, if indirect, proof of its relationship with the active region flux. Overall, the source of the opposing results seems to be the different methods for the definition for active regions: \cite{Kutsenko2017} section the solar surface with a magnetic flux concentration mask on the magnetogram, while \cite{Bose2018} separate active regions from background with a combination of intensity thresholds on the AIA 1600 \AA\, and the 4500 \AA\, images for plage and sunspots respectively.
Nevertheless, assessing the true source of the variations of \Bl is beyond the scope of this paper. We will instead focus on addressing how its behaviours can help us understand stellar activity in the Extreme Precision RV (EPRV) regime and pinpoint stellar characteristics needed for activity modelling.\\

\begin{figure*}
    \centering
    \includegraphics[width=17cm]{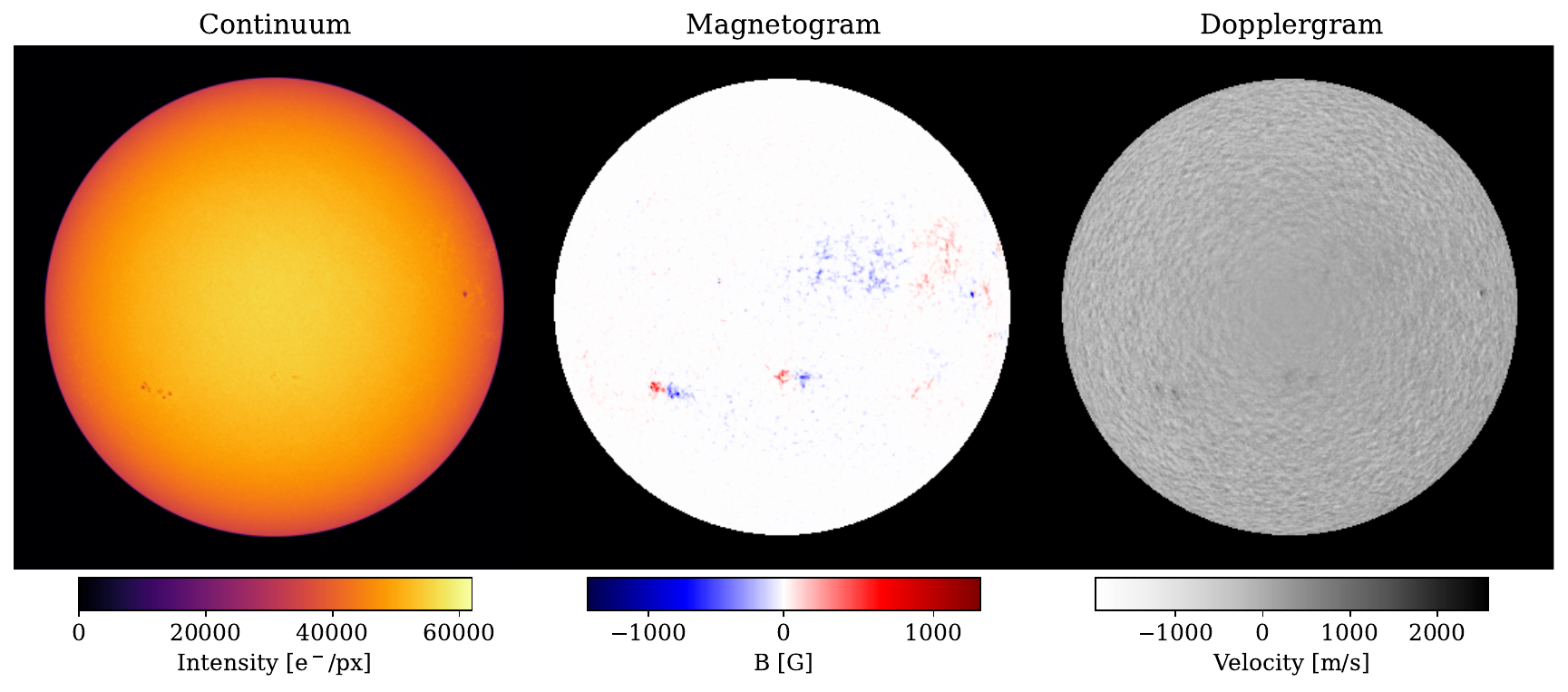}
    \caption{Example SDO/HMI images from 2015-Jul-29. From left to right: the continuum intensity (uncorrected for limb darkening), the line-of-sight magnetogram, and the Dopplergram (corrected for the solar rotation and spacecraft motion).}
    \label{fig:sdo_fulldisk}
\end{figure*}
In this work we use resolved solar observations to extract the mean longitudinal magnetic field of the Sun, and compare it to the radial velocities and the common activity proxies of Sun-as-a-star observations. Since previous works highlighted the effectiveness of \Bl to measure stellar characteristics such as differential rotation \citep{Bruning1991}, the aim of this analysis is to determine whether \Bl can be a useful tracer of stellar activity in Sun-like stars in the context of RV surveys. This paper is organised as follows: we describe the data in Section \ref{sec:data}. We analyse the derived time series to better identify the properties and periodicities of \Bl in Section \ref{sec:full}. Section \ref{sec:sub} covers how we undersampled the data in order to emulate stellar observations, and the tests to assess the ability of the mean longitudinal magnetic field to recover the stellar rotational period and to support RV analysis in a GP regime. We conclude in Section \ref{sec:conclusion}.

\section{Data}
\label{sec:data}
\subsection{HARPS-N Sun-as-a-star Data}
\label{sec:HN}

The HARPS-N solar telescope \citep{Dumusque2015, Phillips2016, CollierCameron2019} is a 7.6-cm achromatic lens which feeds the sunlight to an integrating sphere and through an optical fiber into the High Accuracy Radial-velocity Planet Searcher for the Northern hemisphere spectrograph (\mbox{HARPS-N}: \citealt{Cosentino2012, Cosentino2014}). It is mounted on the Telescopio Nazionale Galileo (TNG) at the Observatorio del Roque de Los Muchachos in La Palma, Spain. Sun-as-a-star spectra are taken continuously throughout the day, with exposure times of 5 minutes in order to average over the solar oscillations. RVs are then extracted using the 2.3.5 version of the ESPRESSO pipeline applied to HARPS-N, the Data Reduction Software \citep[DRS: ][]{Dumusque2021} which computes the cross-correlation function using a G2 stellar mask. From the CCFs, we also calculate the standard activity indicators: the full-width at half maximum and the bisector span. Using the Ca H\&K lines, we also measure the S-index.

Further corrections were applied to the data for them to better represent stellar observations.
The details of these corrections, which we summarise here, are given in \citet{CollierCameron2019} and \citet{Dumusque2021}. First, in order to strip the signal of the Solar System planets, the extracted spectral data are interpolated on the wavelength scale of the heliocentric frame of reference. Next, the effects of differential extinction (noticeable due to the Sun being resolved on the sky) are then removed. The FWHM is corrected for the effects of the Earth's orbital eccentricity and obliquity, which makes the observed projected rotational velocity, and thus the spectral line widths, change over time. Finally, the S-index is corrected for ghosts on the CCD.

Some of our observations will be affected by clouds or other bad weather. To select the best data, we apply strict cuts. The first cut is based on a data quality factor, $Q_{\rm f}$, which is computed using a mixture-model where $Q_{\rm f}=0$ indicates the worst affected data, and $Q_{\rm f}=1$ the data not affected by clouds \citep{CollierCameron2019, AlMoulla2023}. We use only data where $Q_{\rm f}>0.99$. We make a second cut directly using a metric of the exposure meter: the ratio, $R$, of the maximum and mean counts for every observation.
Given the large flux of the solar light, the slight variations and delays in shutter speed of the HARPS-N instrument that are negligible for stellar observations result in a variation in total flux between exposures. This gives a Gaussian distribution in values of $R$ over the full time series. We thus fit the distribution of $R$ with a Gaussian and remove all data where $R$ is higher than the mean plus three standard deviations ($R=1.5$). As a last conservative cut, we perform a 5-$\sigma$ clipping on the remaining RVs. 

In total we considered 64,332 data points from BJD 2457232.873 (2015-Jul-29) to BJD 2459449.104 (2021-Aug-22). All the considered time series are shown in Fig. \ref{fig:HN}.

\begin{figure*}
    \centering
    \includegraphics[width=15cm]{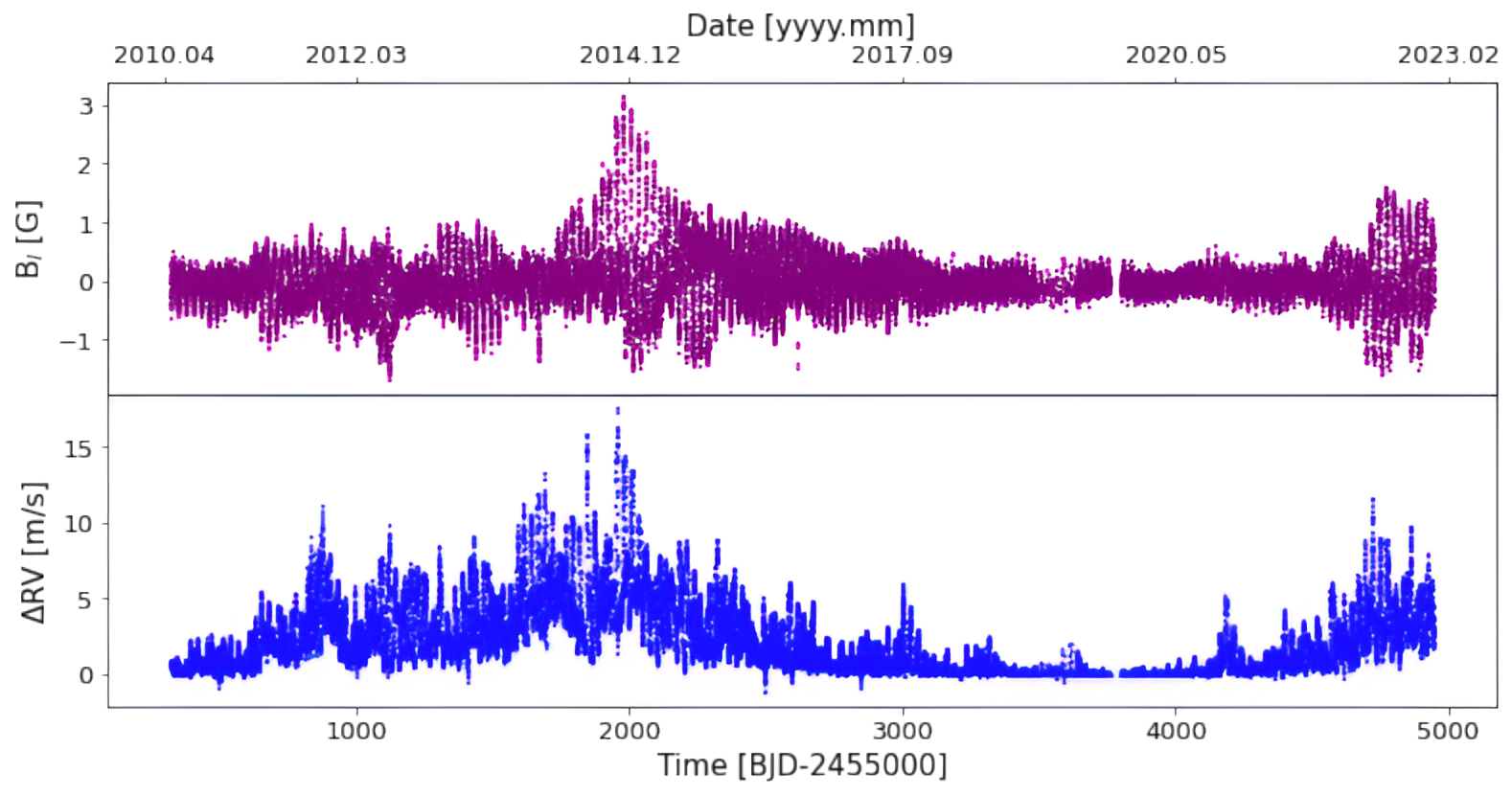}
    \caption{SDO/HMI-derived mean longitudinal magnetic field on the top, and the model radial velocities on the bottom. Uncertainties are not included as they would be too small to be visible.}
    \label{fig:sdo}
\end{figure*}
\subsection{SDO/HMI resolved-Sun images}
\label{sec:sdo}
The Solar Dynamics Observatory (SDO: \citealt{Pesnell2012}) was launched in late 2010 by  NASA's Living With a Star Program, a program designed to understand the causes of solar variability and its impacts on Earth. The aim of the SDO was specifically to study the solar atmosphere on short timescales over many wavelengths. One of its three scientific instruments, the Helioseismic and Magnetic Imager (HMI: \citealt{Schou2012, Scherrer2012}) has been taking continuous full-disc observations of the solar surface with its two cameras of 4096x4096 pixels nearly without interruption since mid-2010. The SDO/HMI instrument has near single-granule resolution \citep{Schou2012,Pesnell2012}. It takes polarised filtergrams of the visible solar disc in two polarisation states by measuring six wavelengths centred in the 6173.3\AA\, neutral Fe I line \citep{Couvidat2016}. Observations are taken every 45 seconds, as well as compiled in 12-minutes (720s) integrated exposures. These filtergrams are then reduced with two main pipelines: the Line of Sight Pipeline and the Vector Pipeline (for more information refer to \citealt{Couvidat2016} and \citealt{Hoeksema2014}). In summary, the six observed wavelengths are fitted with a Gaussian profile to calculate the observable characteristics of the solar surface, such as continuum intensity, photospheric Doppler velocity and magnetic field via Stokes profiles. While the fitting loses any line asymmetry generated within the pixel area, larger processes are preserved.

In this work, we use the 720-second integrated SDO/HMI exposures of the continuum photometric intensity, the Dopplergrams, and the magnetograms reduced with the Vector Pipeline. An example is shown in Fig. \ref{fig:sdo_fulldisk}. While the telescope produces near continuous observations, we choose a cadence of four hours, yielding six images per 24-hour period and 31,755 images spanning nearly 13 years from BJD 2455318 (2010-May-1) to BJD 2459945 (2022-Dec-31).

\subsubsection{Estimating the full-disc solar longitudinal magnetic field and radial velocities} \label{sec:estimate}
While the data is corrected to account for most instrumental effects, long baseline analysis of the solar Doppler velocities was not the original aim of the SDO mission. For this reason long-term stability of the instrument was not prioritised. Therefore, studying the evolution of the solar activity over multiple months or years is not straightforward. To do so a framework was developed that scrambles full-disc images to Sun-as-a-star-like observations coupled with a weighting-based baseline \citep{Haywood2016}. A full pipeline named {\sc SolAster} was introduced by \cite{Ervin2022}. In this work we will briefly cover how time series of RVs and \Bl were generated. For more information refer to \cite{Meunier2010}, \cite{Haywood2016}, \cite{Milbourne2019} and \cite{Haywood2022}.\\

\textbf{Mean longitudinal magnetic field:} In SDO/HMI data the line-of-sight magnetic field, $B_{\rm los}$, is computed for each pixel as the difference of the Doppler velocities observed in two circular polarisations, $V'_{\rm LCP}$ and $V'_{\rm RCP}$:
\begin{equation}
    B_{\rm los} = (V'_{\rm LCP} - V'_{\rm RCP})K_{\rm m},
\end{equation}
in which $K_{\rm m}=0.231405$ for a Land\'e g factor of 2.5. HMI actually directly measures flux density in each pixel, but because a filling factor of one is assumed, a flux density of 1 Mx cm$^{-2}$ is equivalent to a field strength of 1 G \citep{Couvidat2016}. This method is analogous to how the magnetic field is extracted for Magnetic Detection and Imaging (MDI). The 720s version of this variable is computed using selected filtergrams for ten 135s vector fields sequences from Camera 2.

After full-disc foreshortening corrections, we compute the disc-averaged, longitudinal magnetic field of the Sun in each observation by summing the continuum intensity-weighed, line-of-sight magnetic field in each pixel of coordinates $i$ and $j$ on the resolved disc:
\begin{equation}
    B_{\rm l} = \frac{\sum_{ij} B_{\rm los, ij} I_{ij}}{\sum_{ij}I_{ij}},
\end{equation}
in which $I_{ij}$ is the observed, non-flattened continuum intensity in the same pixel. The derived time series is plotted on the top of Fig. \ref{fig:sdo} in purple. The uncertainties of the longitudinal magnetic field at each pixel increase as a function of their position on the disc and distance from the centre, expressed as $\mu$ angle, with them being $\sim$5 G at disc centre and $\sim$8 G at the limbs \citep{Yeo2013}. Even assuming a consistent 8 G noise level, the Poisson-derived uncertainties on the disc-averaged values are incredibly small. Therefore a larger uncertainty will be assumed for the majority of the analysis, as fully addressed in Section \ref{sec:s,data}. The maximum field strength derived for the solar magnetic cycle (Cycle 24 and the beginning of Cycle 25) included in the data is of 3.05 G, which is comparable to the average maximum field derived by the BCool collaboration for G stars of 3.2 G, once again underlining the validity of our comparison.\\

\textbf{Radial-velocity variations:} Given the lack of long-term stability for the Dopplergram data, we compute the disc-averaged RVs starting from a physically motivated model. We define the "quiet" Sun average Doppler velocity as baseline and compute the RVs of each image relative to their respective "quiet" Sun value. We define as active all solar surface larger than 20 ppm (60 Mm$^2$) with absolute foreshortening-corrected magnetic field larger than \mbox{24 G} (see \citealt{Haywood2016} and \citealt{Milbourne2019}). All remaining pixels are then defined "quiet" . We can then compute the baseline value after correcting for the movement of the spacecraft by summing over all "quiet" pixels. With this technique, all RV signals not directly induced by active regions, such as p-modes, granulation or supergranulation, are not included. We are therefore only looking at active region-induced, rotationally-modulated RV variations. These $\Delta$RVs are computed as the linear combination of two active-regions contributions: $\delta$RV$_{\rm phot}$ (the signal generated by active regions breaking the red- and blueshift balance of the rotating disc by enhancing or diminishing the photometric intensity of their area) and $\delta$RV$_{\rm conv}$ (the signal generated by the suppression of convection in magnetic areas yielding to a decrease of the overall convective blueshift on the solar surface). The resulting $\Delta$RVs are plotted in blue on the bottom of Fig. \ref{fig:sdo}.

\section{Full time series analysis: how does \texorpdfstring{$B_{\MakeLowercase{l}}$}{Bl} relate to the RV variations?}
\label{sec:full}
We begin the analysis by assessing the basic properties of the mean longitudinal magnetic field compared to the other derived time series, in order to investigate the behaviour of \Bl over the Solar magnetic cycle.
With this study we aim to answer the following questions: Is \Bl a direct proxy of activity-induced RV variations? How does \Bl change with activity and does it behave like the RVs? What information can we extract from analysing \Bl that cannot be derived from other common indicators? Can we use \Bl to trace and model the solar magnetic cycle?
\begin{figure}
    \centering
    \includegraphics[width=\columnwidth]{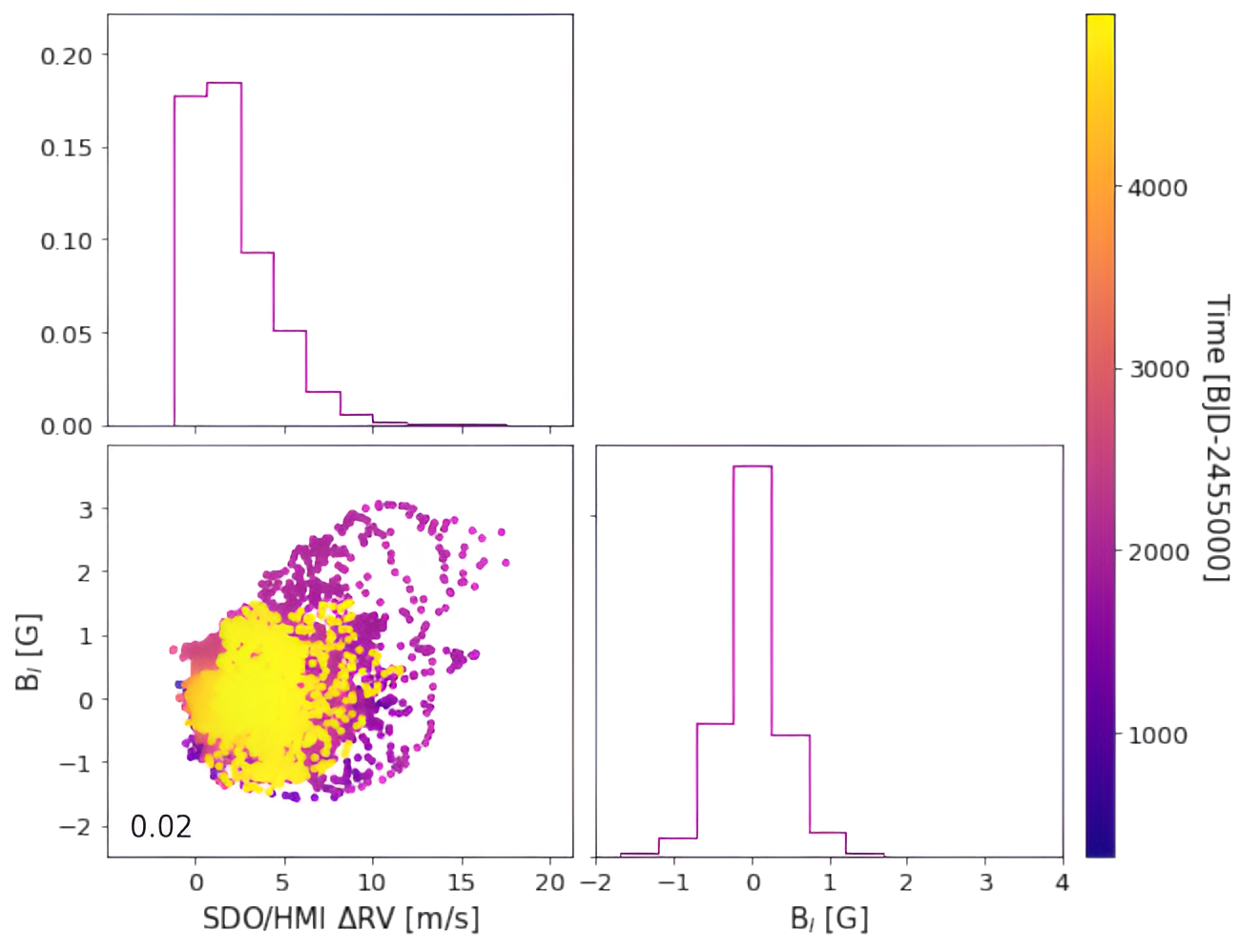}
    \caption{Correlation plot between the SDO/HMI-derived radial velocities and the mean longitudinal magnetic field. The colour indicates the julian date of each datapoint. The computed Spearman rank correlation factor is also included.}
    \label{fig:sdo_corr}
\end{figure}
\begin{figure}
    \centering
    \includegraphics[width=\columnwidth]{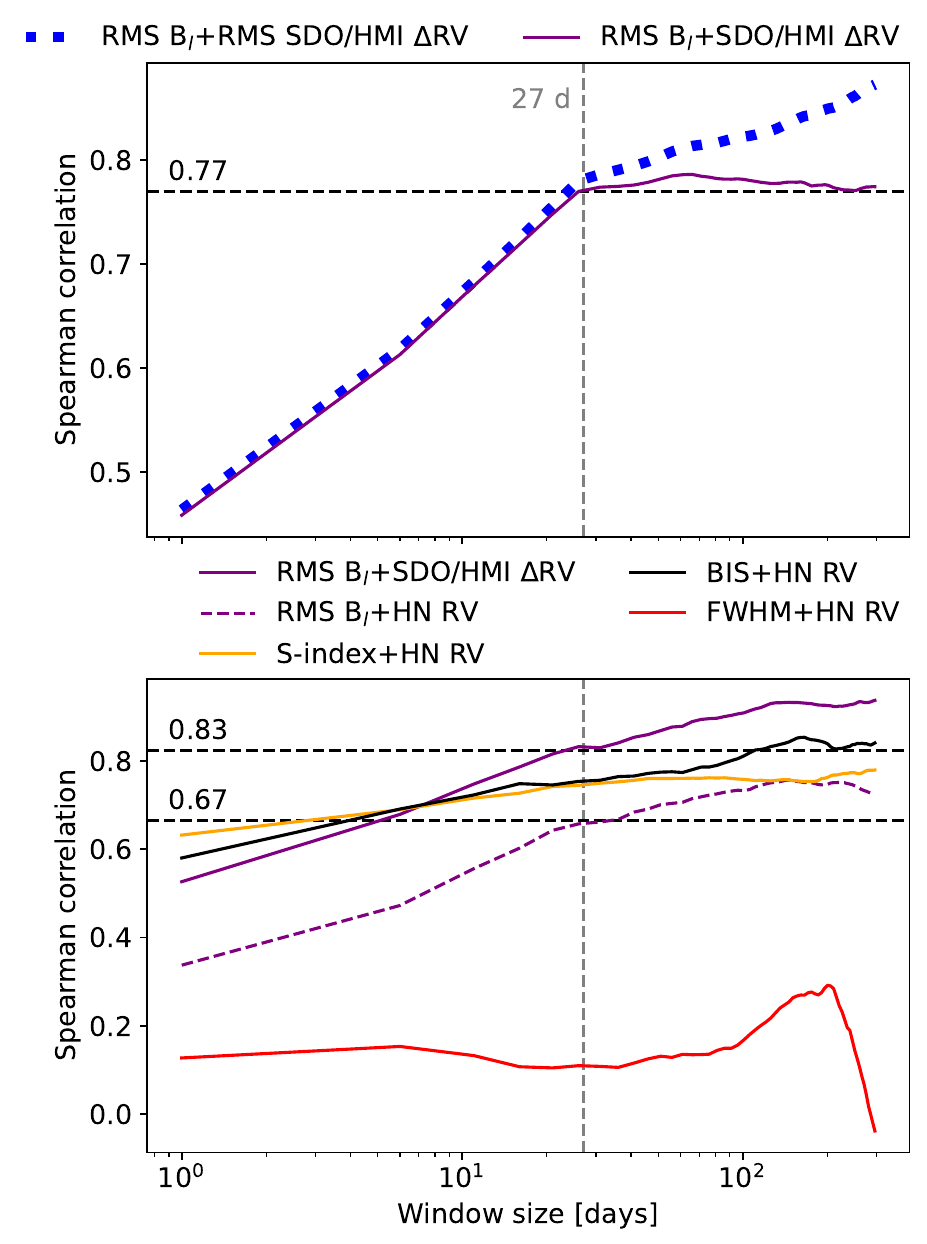}
    \caption{Spearman rank correlation coefficient between two time series against the size of the window (in days) used to smooth the signal (in logarithmic scale). \textit{Top:} correlation between the RMS of \Bl and the RMS of the $\Delta$RVs in blue dotted line, and between the RMS of \Bl and the time-aware mean of $\Delta$RVs as a purple solid line. All considered time series are derived from SDO/HMI data and include all available observations. \textit{Bottom:} Spearman correlation coefficient with varying window size. The time series considered have been matched following the method in Section \ref{sec:match}. Colours represent, in order, the correlations between the RMS of \Bl and  the time-aware mean of SDO/HMI $\Delta$RVs (solid purple), between the RMS of  \Bl and time-aware mean of the HARPS-N RVs (dashed purple), between the time-aware mean of the bisector span (black), the S-index (yellow), and the FWHM (red) with the HARPS-N RVs. The smoothing window equal to a solar rotation period is highlighted with a vertical gray dashed line. Horizontal black dashed lines indicate the correlation coefficient achieved when smoothing over this window.}
    \label{fig:knee}
\end{figure}
\begin{figure}
    \centering
    \includegraphics[width=8cm]{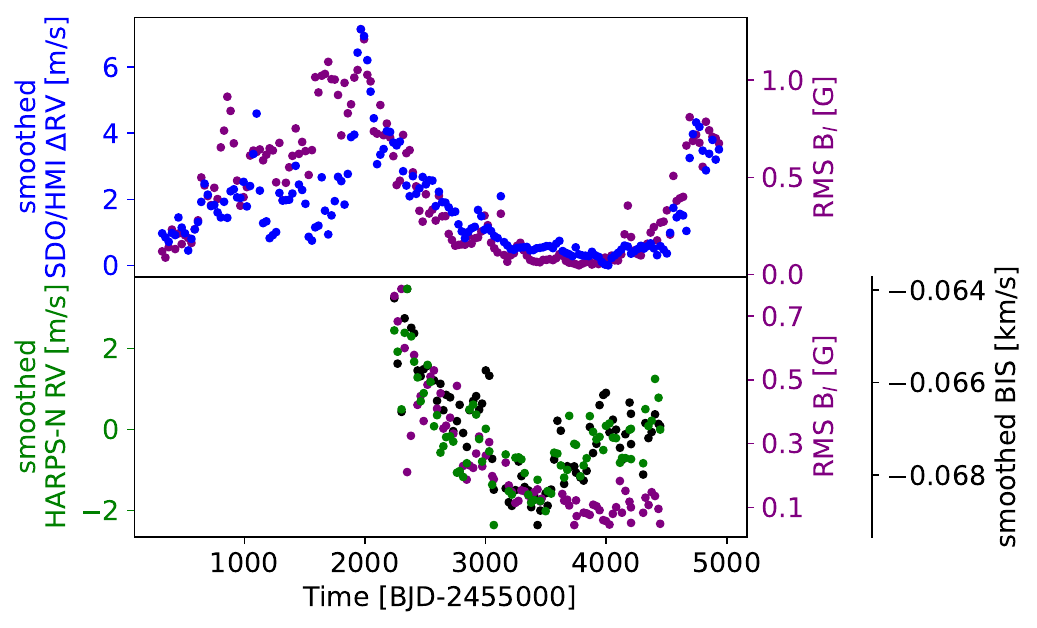}
    \caption{\textit{Top:} time series of the time-aware mean over an averaging window of 27 days of the SDO/HMI $\Delta$RVs in blue. In purple the RMS of \Bl over the same window. \textit{Bottom:} time series of the time-aware mean of the matched HARPS-N RVs in green, of the matched bisector span in black, and the RMS of the matched \Bl in purple over the a window of 27 days.}
    \label{fig:smooth}
\end{figure}

\subsection{Full time series correlation analysis}
\label{sec:full_corr}
To assess whether the mean longitudinal magnetic field can be used to directly map the SDO/HMI rotationally-modulated stellar activity-induced RVs, we compute the Spearman rank-order correlation coefficient of the two time series. When considering all 13 years of data, we calculate a correlation coefficient of 0.02, as shown in Fig. \ref{fig:sdo_corr}, indicating that \Bl does not correlate with the contemporaneous SDO/HMI-derived $\Delta$RVs. We also compute the correlation between the absolute values of \Bl and the $\Delta$RVs. Their Spearman rank correlation coefficient is equal to 0.42, a low moderate correlation.

We compare these results to the correlations calculated between the entirety of the HARPS-N RVs and its activity indicators: 0.54 with the S-index, 0.06 with the FWHM, and 0.52 with the bisector span, as plotted in the Appendix in Fig. \ref{fig:HN_corr}. With the exception of the FWHM, the HARPS-N radial velocities correlate well with the indicators most commonly employed in stellar activity analyses. In particular, a visual inspection of the time series also shows that the BIS and the S-index are sensitive to the long-term trend of the magnetic cycle.
These similarities between the RVs and the activity proxies are at the basis of many mitigation techniques. This good correlation however is not stable in time nor in activity level. In fact, during periods of minima the correlation becomes completely negligible, as we will address in Section \ref{sec:chunks}. This behaviour can be attributed to the fact that these proxies are sensitive to a mixture of different active regions, such as spots, faculae, and network \citep{Cretignier2024}. At low activity other effects not tied to active regions (and therefore not probed with traditional indicators) dominate the stellar variability \citep{Lakeland2024}. Nevertheless, we can now answer the first of our questions: \Bl cannot be used as a direct one-to-one proxy to correct for stellar activity in the radial velocity over all timescales.
The significantly worse correlation (especially when considering all levels of activity) between \Bl and the $\Delta$RV versus the one between the HARPS-N RVs and their proxies is to be expected after a simple visual inspection. As an example, \Bl oscillates between positive and negative values around a mean value of 0.02 G that is stable in time. The mean value of the $\Delta$RVs changes with magnetic cycle phase, going roughly from 2.4 \ms at high activity, to 0.3 \ms during minimum.

We however notice a general trend shared between the two time series through the solar cycle. We postulate that, while direct measurements do not correlate, the root-mean squared scatter (RMS) of \Bl may correlate to the general envelope shape of the RVs, and could therefore be useful information to model the long-term variations due to the magnetic cycle. To test this theory we extract two new time series: we compute the rolling RMS of \Bl over an "averaging window" of a day and the rolling time-aware mean of the SDO/HMI $\Delta$RVs over the same window. The correlation between these new time series improves to 0.48. In order to find the best averaging window size, we repeat the same steps with window lengths between one day and one year. The results of this analysis are plotted in the top panel of Fig. \ref{fig:knee} as a purple solid line. We also include the correlations between the RMS of \Bl and the RMS of the $\Delta$RVs for all the window sizes as a blue dotted line. Both correlations increase steadily until a window size of 27$\pm$1 days reaching a coefficient value of 0.77. At this point, the time series are not mapping the rationally-modulated variations, and are only sensitive to the overarching magnetic activity over the cycle, as shown in the top panel of Fig. \ref{fig:smooth}.
The RMS of \Bl over windows larger than the solar rotation period are able to successfully  map the long-term variations in the $\Delta$RVs. They can therefore be used to correct for the long-term magnetic activity signal via techniques such as contemporaneous fit, or can be selected as training set for a squared exponential kernel in a GP regression framework.
As a simple test, we find the best-fit sine function to the RMS over a window of 27 days of \Bl. We then use the derived parameters to subtract the magnetic cycle long-term trend in the $\Delta$RVs. This very rudimentary method is able to flatted the $\Delta$RVs and reduce their RMS by more than 60\%. 

\begin{figure}
    \centering
    \hspace{-0.8cm}
    \includegraphics[width=\columnwidth]{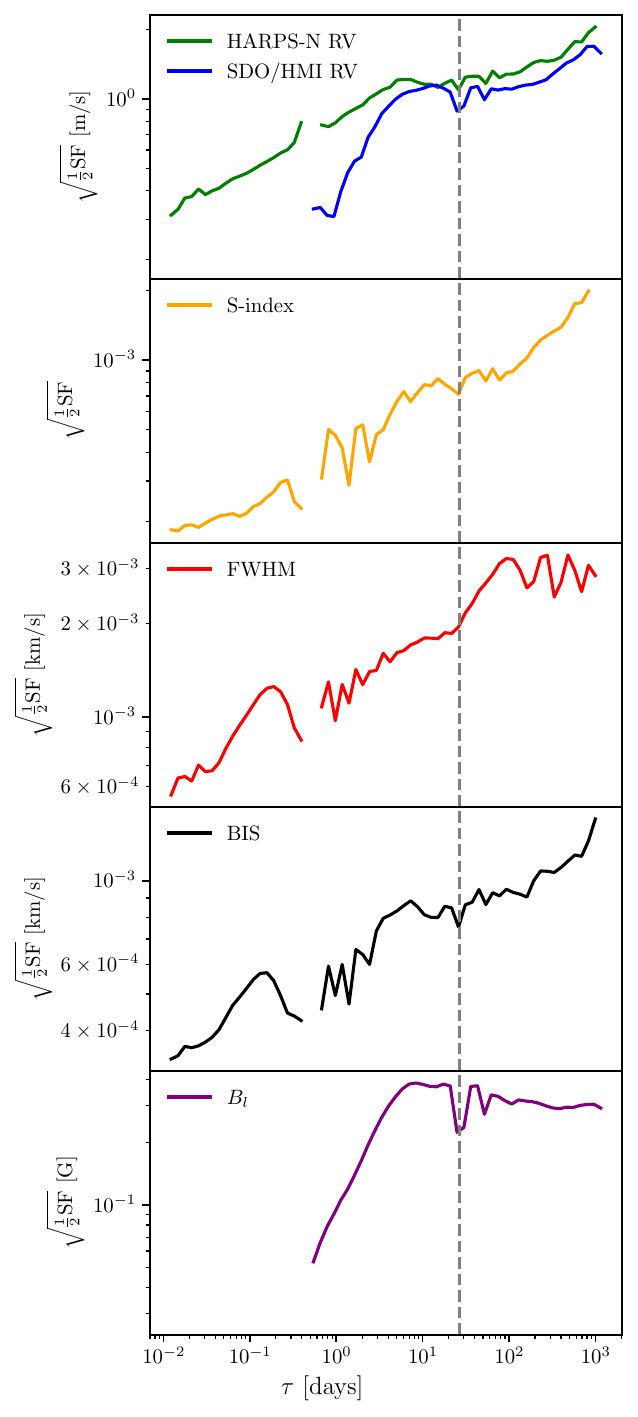}
    \caption{Structure functions of the time series shown in Figs. \ref{fig:HN} and \ref{fig:sdo}. See the main text for more details. From top to bottom: the structure functions for the HARPS-N and SDO/HMI $\Delta$RVs, the S-index, the CCF FWHM, CCF bisector span, and the mean longitudinal magnetic field. The higher cadence of the HARPS-N data is visible in the structure functions as the smaller minimum timescale. Likewise, the diurnal cycle of the ground-based observations gives rise to a gap in the structure function at $\sim$ 0.5 days, since there are no pairs of observations separated by this timescale. The solar rotation period at 27~d is indicated by a grey dashed line.}
    \label{fig:structure_functions}
\end{figure}
\begin{figure*}
    \centering
    \includegraphics[width=15cm]{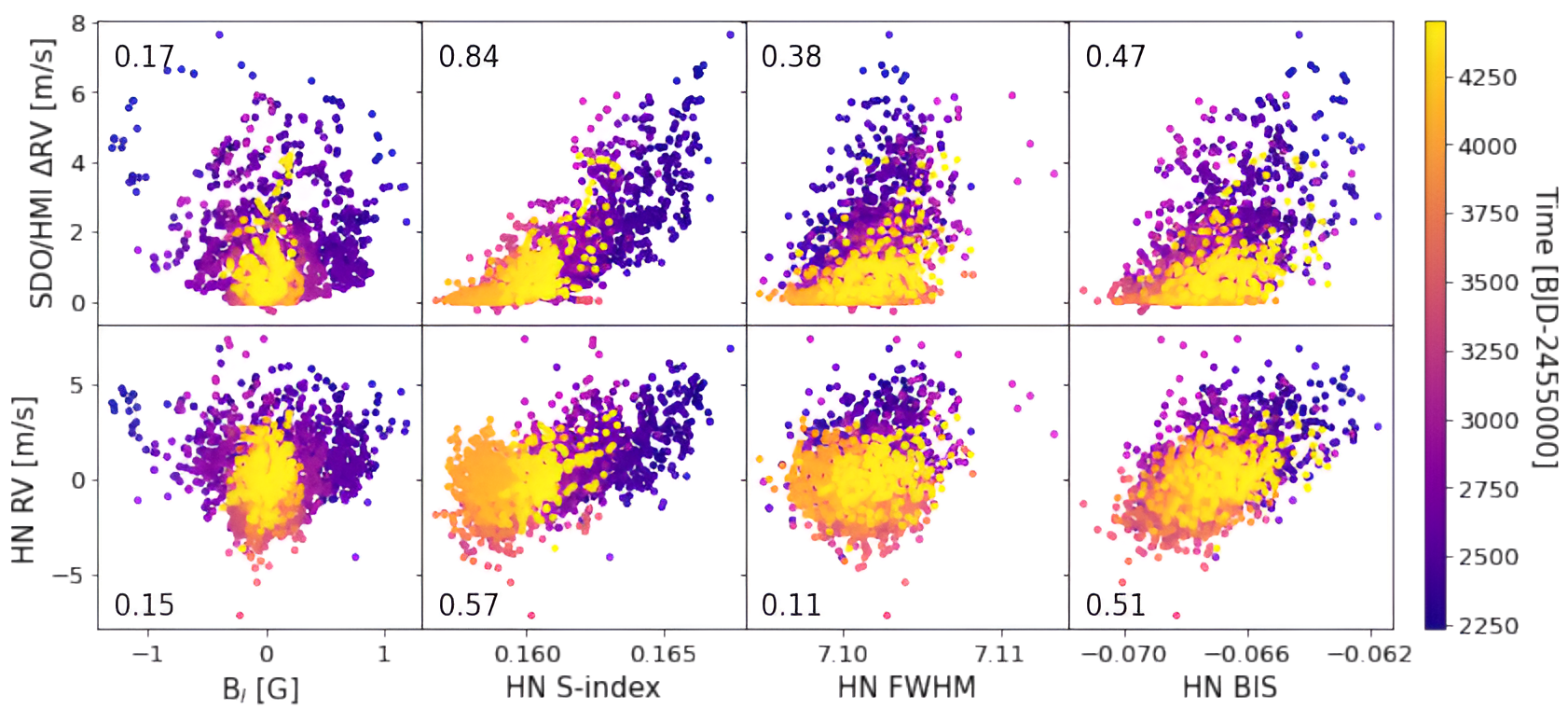}
    \caption{Correlation plots between the matched HARPS-N (here HN) and SDO/HMI time series. The SDO/HMI $\Delta$RVs are in the first row, while the HARPS-N RVs are in the second. From the leftmost to the rightmost column, we plot the mean longitudinal magnetic field, the S-index, the full-width at half-maximum, and the bisector span. The data is colour-coded based on observation time and the Spearman rank correlation coefficients for each set are also included.}
    \label{fig:corr_rv}
\end{figure*}
\subsection{Structure Functions}
\label{sec:full_SF}
We also compute the structure functions of all time series. The structure function measures the variability in a time series at each timescale. It is calculated as
\begin{equation}
    \mathrm{SF}(\tau) = \left\langle \left( f(t) - f(t-\tau)\right)^2\right\rangle,
\end{equation}
where the average is taken over all pairs of observations $f(t)$ separated by time $\tau$. 
\citet{Lakeland2024} show that, for a continuous, uncorrelated signal, $\mathrm{SF}(\tau) = 2 \times {\rm RMS} ^2$. 
We therefore follow the example of \citet{Lakeland2024} and use $\sqrt{\frac{1}{2} \mathrm{SF}}$ to quantify the variability at each timescale, to better draw analogy with the RMS.
We highlight here a few key properties of structure functions; for a more comprehensive review, see \citet{Simonetti1985}, \citet{Sergison2020}, \citet{Lakeland2022}, and references therein.
Firstly, structure functions typically increase with timescale. 
That is to say that two observations taken further apart are likely to be more different than two observations taken closer together.
A notable exception is for a periodic signal, in which observations separated by the period or multiples thereof have similar values. 
In a structure function, such a periodicity manifests as sharp dips.
The second important feature of a structure function is the location of the transition between the region of increasing variability and the plateau. 
This transition occurs once observations separated by greater timescale no longer show more variation. 
The timescale at which this occurs is the characteristic variability timescale of the signal. For timescales longer than this characteristic $\tau$, no additional intrinsic variability is present. In practical terms, to effectively sample a specific source of variability, observations should be taken with time lags within the increasing region of the SF.
In Fig. \ref{fig:structure_functions}, we plot the structure functions of the time series shown in Figs. \ref{fig:HN} and \ref{fig:sdo}. 
To ensure each structure function is well sampled (i.e., with many pairs of observations contributing to each SF calculation), we require at least 50 pairs of observations in each $\tau$ bin.

To allow for direct comparison, we only consider the SDO/HMI data over the overlapping years with HARPS-N (2015 to 2021). 
In Fig. \ref{fig:structure_functions}, the higher cadence of HARPS-N is highlighted by the presence of data at shorter timescales, while the SDO/HMI-derived time series have a minimum $\tau$ of 4 hours. Both RV time series (HARPS-N in green, and SDO/HMI in blue) have similar structure function behaviours. They both increase until a timescale of $\sim$10 days, they then grow at a significantly slower rate (forming plateaus of sorts), to finally start increasing more strongly after 100 days, as the structure functions probe the solar rotation and activity cycle respectively. The RV RMS due to solar activity is of the order of 1 \ms.
A somewhat similar behaviour is shown by \Bl, with an initial increase until $\sim$10 days, and a plateau at a $\sqrt{\frac{1}{2} \mathrm{SF}}$ of $\sim$0.3 G (a low value expected for the extended minimum covered by the considered time series). Note that the structure function shows a slight decrease in this plateau region. 
We explain this behaviour by considering the magnetic cycle. Overall, the signal of \Bl at comparable levels of stellar activity over different cycles is similar. That is to say that there are similarities between variations \Bl at the rise and the decline of Cycle 24 versus the rise of Cycle 25.
The most interesting feature to notice in the structure function of the magnetic field is the significant dip at $\sim$27 days (and a second smaller one at $\sim$55 days). These dips highlight the strong modulation of the time series at the solar rotation period. 
The SDO/HMI-derived radial velocities also show similar dips at one- and two-times the solar rotational period. While the HARPS-N RVs do share this feature, it is much less prominent.
This is because the HARPS-N RVs are calculated using thousands of spectral lines and are sensitive to variability caused by additional physical processes on the Sun beyond those directly linked to magnetic activity, whereas the RVs calculated from SDO/HMI only consider the effect of active regions, which show a stronger rotational modulation.

All three HARPS-N activity indicators have very similar structure functions, with a gradual but consistent increase at all timescales. They all present dips at $\sim$1.5 and 2.5 days, which are not replicated in their radial velocities.
The lack of a plateau region in any of the HARPS-N activity proxies means that no characteristic timescale of the variability can be retrieved: the activity indicators are affected by multiple physical processes all with different timescales. On the other hand, \Bl shows a characteristic timescale of the order of half the solar rotation period, meaning that its behaviour can be sampled with two observations per period.
This analysis therefore highlights the elevated complexity of the signal of the common activity proxies versus the mean longitudinal magnetic field, and is a first proof of the strong rotational modulation of \Bl.

\subsection{Matching the data between HARPS-N and SDO/HMI}
\label{sec:match}
As the timestamps for the HARPS-N and SDO/HMI data are different, it is necessary to match observations of the two time series in order to investigate the relationship between \Bl and the RVs from HARPS-N.
To do this, we interpolate the SDO/HMI data onto the timestamp of the nearest HARPS-N observation if the time between the two is less than one hour.
If the time between an SDO/HMI observation and its closest match in the HARPS-N data set is more than one hour, the data point is omitted.
A justification for this approach is provided in Appendix \ref{sec:appendix_matching}.
The resulting time series has 2,891 data points and includes the diurnal cycle and realistic poor-weather breaks from the HARPS-N data, and the maximum of six observations per 24-hour period of the selected dataset for SDO/HMI.

\subsection{Correlation Analysis}
\label{sec:match_corr}
We then reassess the correlation and recompute the Spearman coefficients between both the HARPS-N RVs and SDO/HMI $\Delta$RVs, with all the considered activity tracers, including \Bl, as shown in Fig. \ref{fig:corr_rv}. As derived previously in Section \ref{sec:full_corr}, \Bl does not correlate with either of the radial velocities. It is interesting to note, however, that correlations between the HARPS-N activity indicators and the RVs derived with the same instrument are lower than (or in the case of the BIS, comparable to) their correlation to the SDO/HMI radial velocities. As mentioned previously, the SDO/HMI $\Delta$RVs are only sensitive to rotationally-modulated active region-induced RV variations, while the HARPS-N RVs are additionally influenced by all other physical processes on the solar surface as well as instrumental systematics. These results highlight why these indicators are not able to successfully map the RV signals imprinted by processes such as granulation or supergranulation. We also replicate the same averaging window study undertaken in Section \ref{sec:full_corr}, as shown in the bottom panel of Fig. \ref{fig:knee}. We compute the correlations between the RMS of the matched \Bl and the time-aware mean of the SDO/HMI $\Delta$RVs with a rolling window size between one day and one year. The results are plotted as a purple solid line. With a window size of roughly the solar rotation period, the two time series reach a strong correlation of 0.83, reconfirming our earlier conclusion. For comparison, we also plot the correlations between the time-aware mean of the matched time series of the HARPS-N activity indices and their RVs. The FWHM (in red) does not map the long-term trend and therefore does not correlate well over all considered windows. On the other hand, the S-index and the BIS (respectively in yellow and black) reach similar correlations of 0.75 at a window size of $\sim$27 days. Thus, the RMS of \Bl over a solar rotation period correlate better to their smoothed $\Delta$RVs than the HARPS-N activity proxies do to the smoothed RVs derived from the same instrument.
As a further test, we also include the correlation with increasing rolling window size between the RMS of the matched \Bl and the matched HARPS-N RVs, plotted as a dashed purple line.  Unlike before, the correlation is lower, with it reaching only 0.67 at the solar rotation. To investigate this behaviour, in the bottom panel of Fig. \ref{fig:smooth}, we plot the matched HARPS-N RVs and BIS time series smoothed over a solar rotation period (in green and black respectively) alongside the RMS of the \Bl over the same window (in purple). It is clear that, while the RMS of \Bl matches the slow general decrease at the end of cycle 24, the time series extracted from SDO/HMI diverge from the ones derived from the HARPS-N spectrograph around BJD 2458500, roughly the start of the extended minimum. The SDO/HMI $\Delta$RVs follow the same shape as the RMS of \Bl, instead of bending back up, as the HARPS-N data do.
This different trend between HARPS-N and SDO/HMI can be caused by a variety of sources, the study of which is above the scope of this work.

\begin{figure}
    \centering
    \includegraphics[width=\columnwidth]{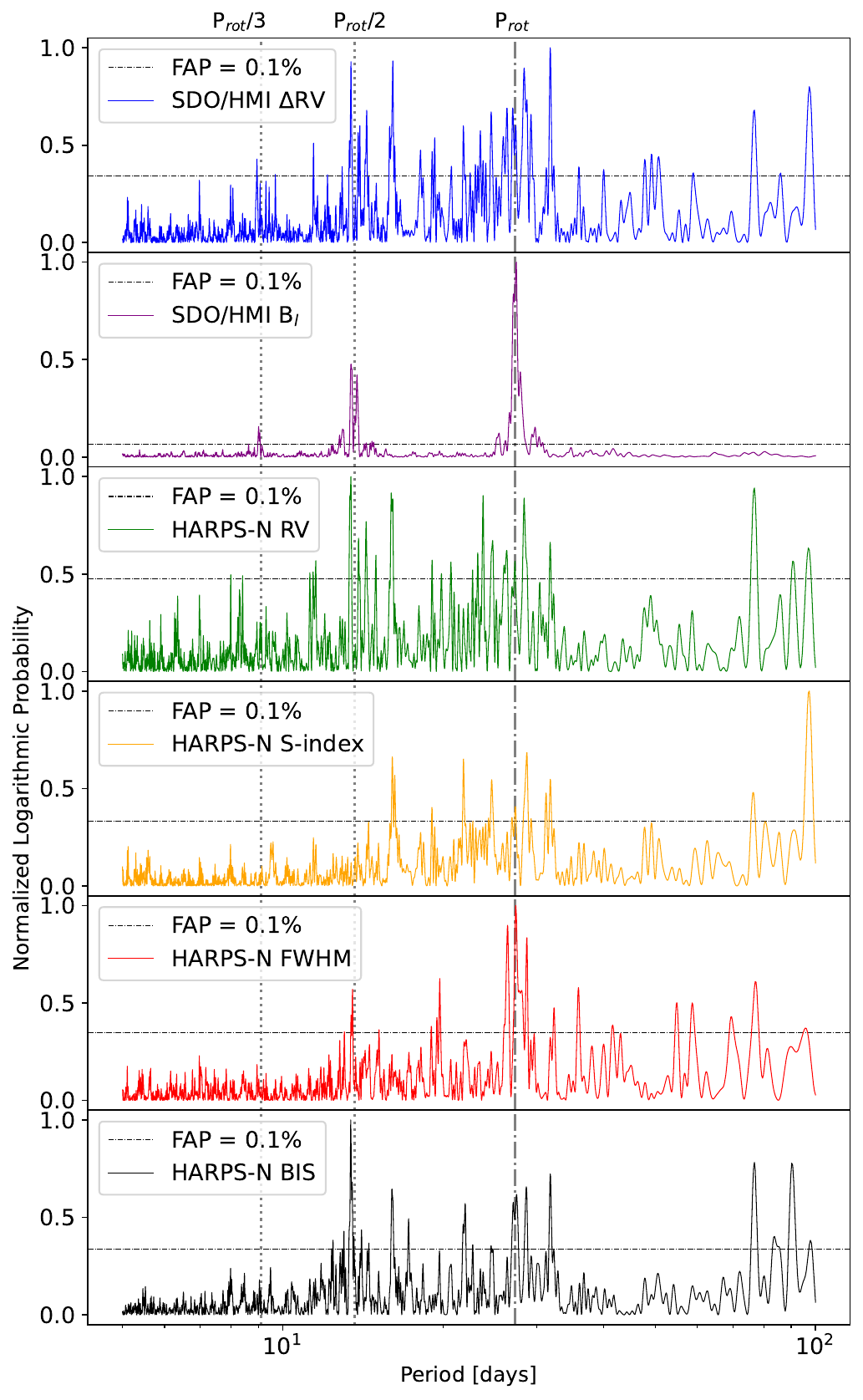}
    \caption{Generalised Lomb-Scargle Periodogram of the matched time series. On the x-axis the period in days, on the y-axis the normalised logarithmic Zechmeister-K\"{u}rster power (or probability). From top to bottom, the matched time series of SDO/HMI $\Delta$RVs, and mean longitudinal magnetic field, HARPS-N RVs, S-index, FWHM, and bisector span. The Carrington Solar rotation period is indicated by a gray dash-dotted line. Its half- and third-period harmonics are also highlighted by dotted lines. The False Alarm Probability (FAP) equal to 0.1\% are included as dashed gray horizontal lines.}
    \label{fig:periodograms}
\end{figure}
\subsection{Periodogram Analysis} 
\label{sec:match_period}
We compute the Generalised Lomb-Scargle periodograms (GLS: \citealt{Zech2009}) of all the time series for the complete and the matched datasets. Both produce similar results. For this analysis, we focus on signals with periods smaller than 100 days, as longer magnetic cycle periodicity would not be reliably picked up with the available baseline. In particular, we are interested in assessing the ability of \Bl to systematically recover the solar rotation period. For these reasons, we remove all long-term signals with a low-pass filter.
Via this comparison we are also able to confirm that no significant periodic signal is introduced in the data-matching step by interpolating the SDO/HMI data on the HARPS-N timestamps.
In this work, we only include the periodograms for the matched datasets and plot them in Fig. \ref{fig:periodograms}.  In the figure, the Carrington solar rotational period of $P_{\rm rot}=$27.2753 days as seen from the Earth is highlighted as a reference with a gray dash-dotted line. We also include the first and second harmonic of the Carrington period as dotted lines. By quick visual inspection, it is clear that the mean longitudinal magnetic field strongly outperforms all activity indicators in finding the expected rotational period. The only other relevant peaks in the periodogram of \Bl are generated by the first and second harmonics of $P_{\rm rot}$. This behaviour has been noted before for other \Bl measurements \citep[e.g.,][]{Kotov1983, Grigorev1987,Obridko1992} and is similar to the results obtained by \cite{Xie2017} via wavelet transformation.

Both RVs are slightly more sensitive to half rotational period than the full one, although they present wide forests of peaks at $P_{\rm rot}$ and $P_{\rm rot}/2$ both. The S-index has its most significant peak around 100 days, followed by one at $\sim$29 days. Even considering all peaks above the 0.1\% False Alarm Probability (FAP) level, the S-index does not reliably recover the solar rotation period. The FWHM is the most sensitive out of the HARPS-N proxies to $P_{\rm rot}$, with a forest of peaks centred in $\sim$29 days. It also shows peaks at $\sim P_{\rm rot}/2$, as well as $\sim$19 days. The periodogram of the BIS is nicely peaked around $P_{\rm rot}/2$, with some signal around $P_{\rm rot}$, as well as $\sim$32 and 22 days. The periodograms of all time series excluding \Bl are complex at high frequencies, and have power at longer periods. Finally, the true solar $P_{\rm rot}$ cannot be recovered to a reasonable level of accuracy or precision from this analysis.
In contrast, the periodogram of the mean longitudinal magnetic field is much simpler and does not present any significant power at long periods. As mentioned previously all the power is concentrated at $P_{\rm rot}$ and its harmonics. In fact, given the formulation of periodograms, we should not expect any power at low frequencies. Periodograms fit sinusoidal curves to the data for all periods and assess the goodness of the fit. Long-term effects can be fit by a sine curve in the radial velocities and its common proxies, but they behave differently in \Bl.
As an example, the magnetic cycle imprints on the RVs a general increase in their mean value over time of maximum and a decrease over times of minima. On the contrary, the average value of \Bl stays constant in time. The magnetic cycle only affects the amplitude of the oscillations, not their mid-point, meaning that they cannot be fit by a long-period sine curve. This effect yields a much simpler periodogram.
We can overall conclude that a Fourier analysis of \Bl is significantly more sensitive to $P_{\rm rot}$ and allows for a much more precise and accurate identification of the solar rotation period than all other analysed time series.\\

\begin{figure}
    \centering
    \includegraphics[width=\columnwidth]{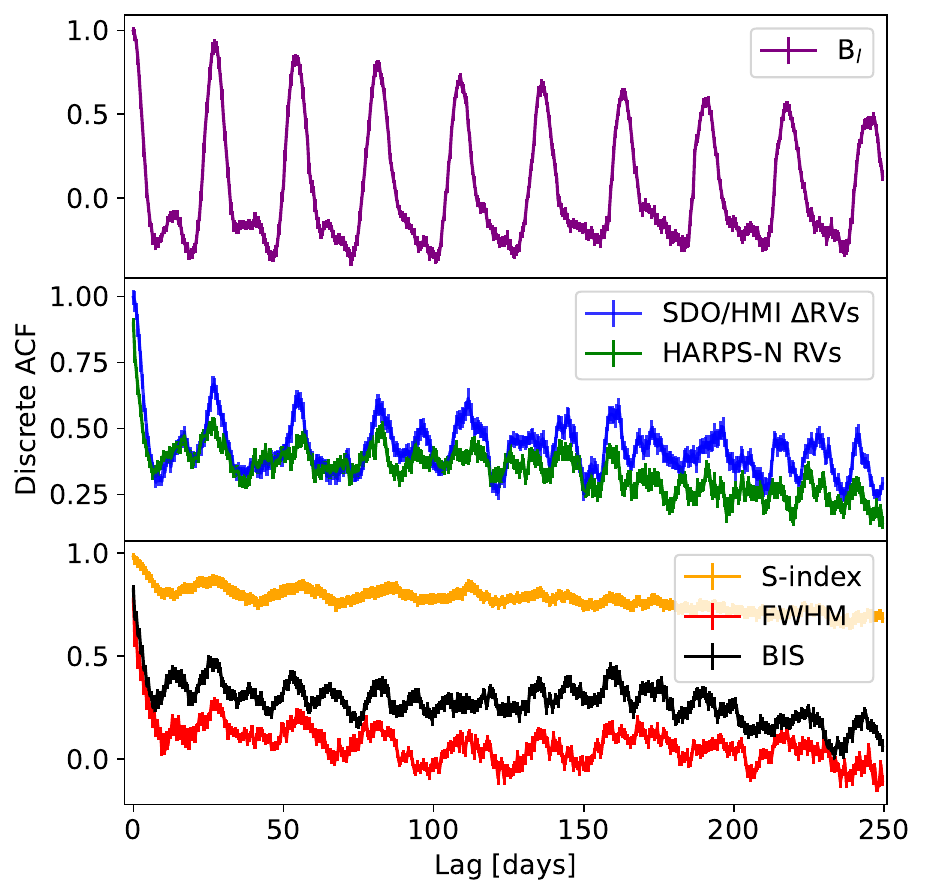}
    \caption{Autocorrelation function over a lag window of 250 days of the mean longitudinal magnetic field (top row in purple), the SDO/HMI and HARPS-N RVs (middle row in blue and green respectively), and the HARPS-N activity proxies S-index, full-width at half-maximum and bisector span (bottom row in orange, red and black). Uncertainties are included as errorbars.}
    \label{fig:acf}
\end{figure}
\subsection{Autocorrelation Function Analysis}
\label{sec:match_AFC}
Another way of isolating the rotational period is to compute the autocorrelation function of the time series \citep{Giles2017,CollierCameron2019}. An autocorrelation analysis measures the relationship between observations at different points in time, and can therefore isolate patterns over the time series.
We use the method developed by \cite{Edelson1988} and updated in \cite{Robertson2015} to compute the autocorrelation function (ACF) for unevenly sampled datasets. In very simple terms, we "slide" in time the data and compute how well it correlates to its original version via Pearson rank-order correlation coefficient. Our code is adapted from {\sc pydcg}\footnote{Available at: \url{https://github.com/astronomerdamo/pydcf}}. Assuming significant rotational modulation, the solar rotation period can be extracted as the lag between each major peak in the ACF. We obtain the ACF for all matched time series, as shown in Fig. \ref{fig:acf}. As in Section \ref{sec:match_period}, \Bl is especially good at recovering the rotational period of the Sun, and its periodic signal stays strong and clear over multiple rotations.  We compute the half-life of the autocorrelation to be 2.74$\pm$0.02 days.
While not wholly insensitive to the rotation period in this analysis, the HARPS-N RVs and the other proxies do not show as clear or well-peaked signals. As expected, the SDO/HMI $\Delta$RVs present a smoother ACF than the HARPS-N ones, as they are derived with a model that considers only rotationally-modulated components.

\begin{figure}
    \centering
    \includegraphics[width=\columnwidth]{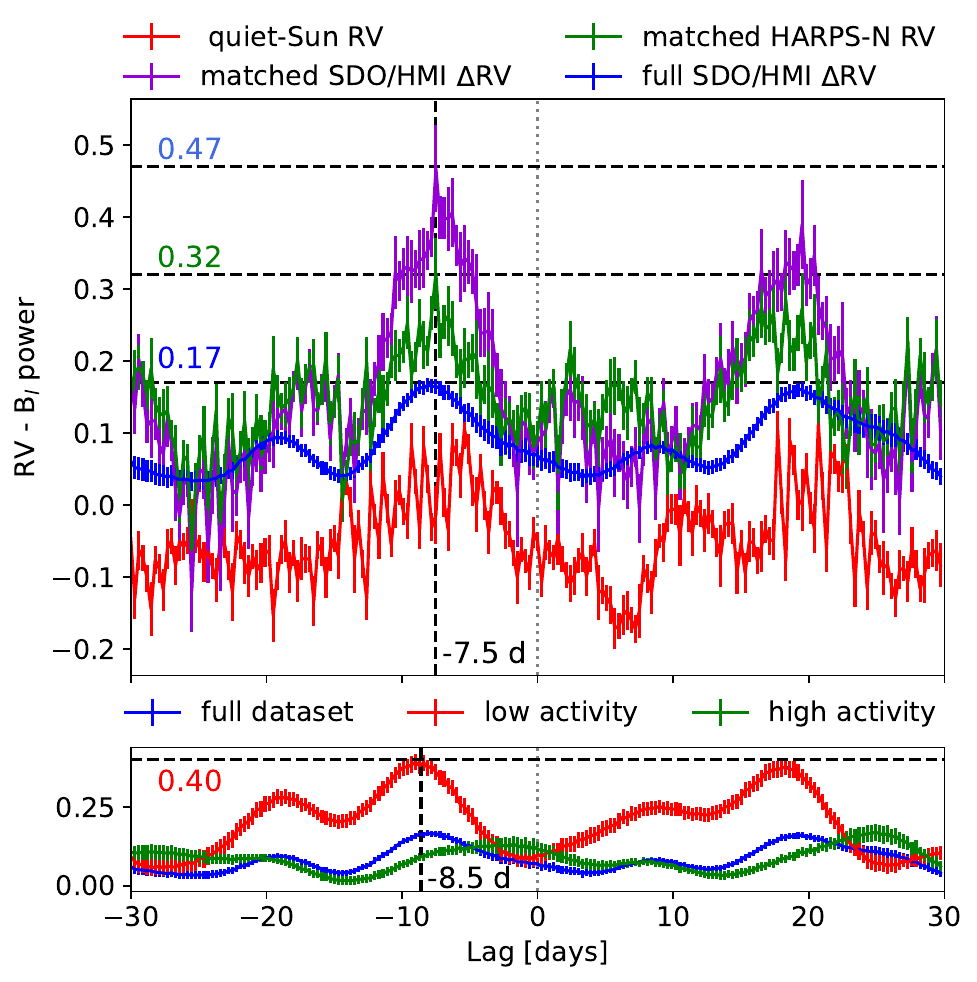}
    \caption{\textit{Top:} lag plot between \Bl and the RV time series. The lag against the matched SDO/HMI $\Delta$RVs are plotted in pale blue, while the lag against the HARPS-N matched RVs are plotted in green. We include the lag between \Bl and the full SDO/HMI $\Delta$RVs time series in dark blue. In red we plot the lag between the matched \Bl and the "quiet-Sun" RVs, computed as the subtraction between the matched HARPS-N RVs and the active regions-derived $\Delta$RVs from SDO/HMI. On the y-axis is the Pearson rank correlation coefficient computed between \Bl and the time shifted RVs. Uncertainties on the power are included as errorbars. The best correlation achieved and the best-fit lag are highlighted by black dashed lines. The 0 lag is highlighted with a gray dotted vertical line.
    \textit{Bottom:} lag plot between \Bl and SDO/HMI $\Delta$RVs. In blue, as in the top panel, the full SDO/HMI dataset, in red the low activity section of the same RVs (2015-Dec to 2021-Jan), in green the high activity section of the RVs (2010-May to 2015-Nov). The best fit lag of the low activity RVs and its respective correlation are highlighted with a black dashed lines. The 0 days lag is identified by a vertical dotted gray line.}
    \label{fig:lag}
\end{figure}

\begin{figure*}
    \centering
    \includegraphics[width=15cm]{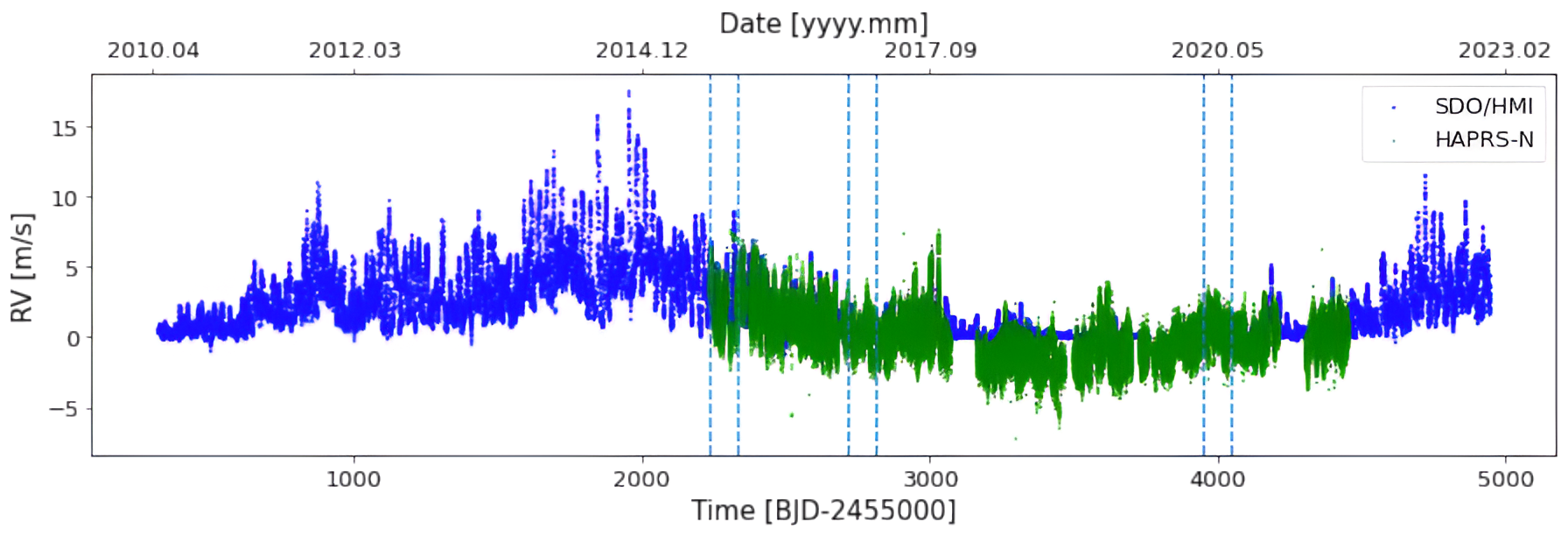}
    \caption{The selected stellar season-like 100 days chunks for the analysis in Section \ref{sec:sub} are shown by vertical dashed lines. The SDO/HMI radial velocities are plotted in the darker blue, and the HARPS-N RVs in the lighter green. Uncertainties are included but may be too small to be visible.}
    \label{fig:chunks_years}
\end{figure*}
\subsection{Lag Analysis}
\label{sec:match_lag}
Previous works have proposed or attempted to constrain the presence of time lags between the stellar activity proxies and the RVs \citep[][Mortier et al. in prep]{Boisse2011,Santos2014,CollierCameron2019,Costes2021}. We therefore also investigate the presence of any significant lag between the mean longitudinal magnetic field and both radial-velocity time series. We use the same method as described in the previous section, this time computing the correlation between two different time series and shifting in time one with respect to the other. We limit our investigation to lag values between $-30$ and 30 days. The results of this analysis are plotted in the top panel of Fig. \ref{fig:lag}. We first examine the cross-correlation function between \Bl and the total SDO/HMI $\Delta$RVs, plotted in dark blue. When considering the entirety of the available 13 years of data, no significant lag can be found. A best correlation of 0.17 is recorded at $\sim-7$ days. As a next step we compute the cross-correlation functions between time-matched datasets. Starting from the matched SDO/HMI $\Delta$RVs, we find a lag of $-7.5\pm0.5$ days with a correlation coefficient of 0.47$\pm$0.05. Similarly, the most probable lag between the matched \Bl and HARPS-N RVs is found at $-7.4\pm0.5$ days with a correlation coefficient of 0.32$\pm$0.05.  We also note second possible peaks for all  RV time series at $\sim$20 days. We interpret this as the repetition of the same lag in the "next" rotational period (assuming a $P_{\rm rot}$$\sim$27 days). The derived best lag is comparable to a fourth of the solar rotation, or roughly the difference between disk centre and limb. We note that, with the same time sampling and baseline, the SDO/HMI $\Delta$RVs reach a higher correlation than the HARPS-N RVs for the same lag. In order to test whether this possible lag is driven by the presence of active regions, we also compute the cross-correlation between the matched \Bl and the "quiet-Sun" RVs. This last time series is computed as the subtraction between the HARPS-N RVs (expected to include all processes) and the SDO/HMI $\Delta$RVs (which only include active region-induced effects). It represents the RV variations caused by all physical processes on the Sun that are not directly tied to either the flux imbalance or the suppression of convective blueshift generated by the presence of large active regions. This method is justified in \cite{Lakeland2024}.  No significant lag can be extracted between \Bl and the quiet RVs. These results point to the conclusion that active regions, such as spots and faculae, are the driving force behind the possible lag between \Bl and the RVs. To further investigate, we also plot in red and green respectively in the bottom panel of Fig. \ref{fig:lag} the cross-correlation function between \Bl and the SDO/HMI $\Delta$RVs during high activity (when the active region filling factors are maximised) and low activity (when active regions are few and far in between). Only the low activity $\Delta$RVs show a clear lag with \Bl at $-8.5\pm0.5$ days with a 0.40$\pm$0.05 peak correlation coefficient. The cross-correlation also peaks at $\sim$20 days, but differently from before, it also presents somewhat significant peaks at  $\sim$9 days (and the related rotation peak at $\sim-19$ days). The high activity cross-correlation is as flat as the one between \Bl and the full $\Delta$RVs (also re-plotted in the bottom panel for comparison). These results seem to oppose our earlier conclusion. However, it is important to note that while at low activity the Sun does develop substantially less spots and faculae than during maximum, the surface is never fully bereft of them. In fact, even though the model to compute the SDO/HMI $\Delta$RVs only considers the direct effects of large active regions, there is still some variability during solar minimum. We can therefore explain these results as follows: at high activity the larger number of active regions allows their longitudinal distribution to be significantly more even over the solar disc. Their contributions to a lag may therefore be "smoothed" away. On the other hand, during minimum active regions can more easily be approximated to a single cluster. This yields a "simpler" signal and the lag can be more successfully recovered. Further analysis is required to truly understand the source and the reason behind this lag.
Nevertheless, this best derived lag between \Bl and the RVs does not reach the necessary correlation threshold of 0.5 to be considered significant. In fact, we are able to prove that the best-fit lag is not constant in time. To confirm this, we divide the SDO/HMI data in rolling 200 days chunks and find the cross-correlation coefficient at the best lag for each section. The computed lags for the 31,555 data chunks range from $-8$ days to 5 days. Roughly 80\% of best-fit lags have correlation coefficients under 0.4. We retrieve no significant trends with time. The $-7.5$ days result is only recovered from the distributions of all the best-fit lags when considering only results with correlation coefficient above 0.4.

\begin{figure*}
    \centering
    \includegraphics[width=15cm]{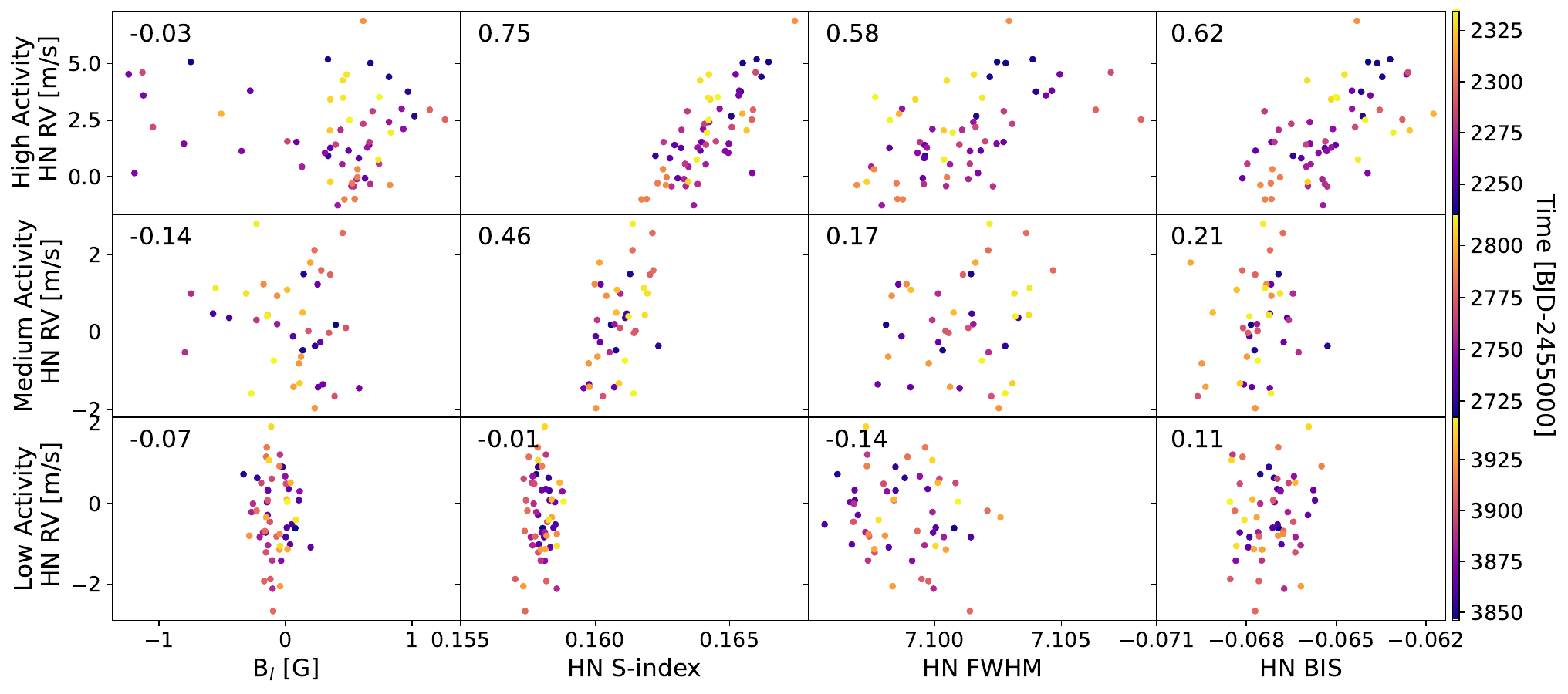}
    \caption{Correlation plots of the (from top to bottom) high, medium and low activity data selection of the HARPS-N (here HN) radial velocities against the considered activity proxies. The data is colour-coded based on date. The Spearman rank-order correlation coefficient of each pair is also included.}
    \label{fig:chunk_corr}
\end{figure*}
Looking back to Fig. \ref{fig:sdo}, a general visual inspection does point to an overarching possible longer time lag between \Bl and the SDO/HMI $\Delta$RVs, especially for the years 2013 to 2015, when the activity is at the highest. This behaviour has been noticed and investigated before: \cite{Sheeley2015} show that during most observed magnetic cycles the solar large scale field undergoes sudden rejuvenation only after the Sun has reached its maximum. They in fact state that a significant increase in the absolute \Bl marks the start of the declining phase of the cycle. Therefore, this increase in amplitude is not directly tied to the increase of solar photospheric activity (measured by the number of sunspots and other active regions, as the maxima of the solar cycle is normally defined). It is instead caused by the longitudinal distribution of sources of flux, in particular a specific arrangement that (together with contribution from the axisymmetric element) reinforces the equatorial dipole component of the magnetic field.
Since the migration and emerging patterns of active regions are expressions of the magnetic cycle of the Sun \citep{Hathaway2015}, the rejuvenation of \Bl is inherently tied with the stage of the cycle.
Overall, this effect means that the increase in \Bl while correlated to the magnetic cycle is not symptomatic of the same process as the increase of variability in the radial velocities (which is primarily dependent on the filling factor of the active regions). The time delay of the maximum amplitudes of the mean magnetic field is therefore explained by the time required by the larger amount of active regions to migrate inward \citep{Huang2017} and emerge in the "correct" arrangement. The length of this "lag" is not fully constrained yet, given the long baselines required to have enough data for a proper statistical approach, but it can be approximated to be of the order of months to a year.
This significant correlation between the pattern of emergence of flux and the value of \Bl can also inform us about the physical locations of the largest active regions on the surface of the star. In fact, \cite{Wang2011} find that the increased bias towards larger positive \Bl values during 2014 was generated by a north-south asymmetry in the distribution of flux emergence. In the Sun, poleward surface flows maintain a north-south asymmetry in the photospheric field, which in turn generates an asymmetric quadrupole component. This means that at times, one sector of polarity can dominate over the other at the solar equator. The overall sign bias of the oscillations of \Bl is therefore dictated by the leading polarity in the more active hemisphere.
In this case the wide positive amplitudes were induced by greater sunspot activity in the southern hemisphere of the Sun, as the southern wing polarity for Cycle 24 was positive \citep[e.g.][]{Norton2023}. Differently from RVs, the mean value of \Bl and its evolution with time inform us about the leading polarity of the active regions, and in cases in which the magnetic field is better understood, they inform us about the hemispheric positions of the active regions.
In time series of stellar observations, this information could also be employed as further constraints in Zeeman Doppler Imaging \citep{Semel1989,Brown1991,Piskunov2002}.

\section{Stellar-like Observations: can we use \texorpdfstring{$B_{\MakeLowercase{l}}$}{Bl} to measure \texorpdfstring{$P_{\rm \MakeLowercase{rot}}$}{Prot}?}
\label{sec:sub}
In Section \ref{sec:full} we have demonstrated that with high cadence and long baseline, the mean longitudinal magnetic field is an effective period detector due to the strength and the simplicity of its signal. 
However, the value of a good activity tracer is its ability to inform us about the stellar variability successfully over much shorter timescales.
Is \Bl as good as a rotational period detector with larger uncertainties and with significantly less data, as is the case with most stellar datasets? Can \Bl be relied on over all levels of magnetic activity, or will it fail at low activity, as do most of the other common activity proxies?
We therefore test the mean longitudinal magnetic field as a "stellar" activity tracer. We limit our data to the average stellar season length, 100 days. We also select three chunks of data over the available years in order to test the effectiveness of \Bl over multiple phases of the solar magnetic cycle, at highest, medium and lowest activity available in the HARPS-N dataset.

\subsection{Choosing a Realistic Stellar-like Cadence and Precision}
\label{sec:s,data}
In order to represent a typical observational schedule for a star in the context of exoplanet detection, we pass through a second data selection process. A typical cadence for stars is maximum of an observation a night. We therefore select only one observation taken each day of data. We do not average all datapoints to daily bins, as that would get rid of effects such as granulation and it would not be representative of the type of observations undertaken for stars.
We instead randomly select one observation over each 24 hour window.
In this analysis we do not account for the difference of integration time per exposure. The 5 minutes exposure length of HARPS-N solar data is long enough to average out p-modes, and all other physical process that can significantly influence the RV variations (e.g., supergranulation) have baselines longer than the average exposure time of stellar observations. 
We select three 100-day chunks over three stages of stellar activity. A high stellar activity case is selected for BJD 2457235 to 2457335 (2015-Ju-31 to 2015-Nov-8), close to the start of the HARPS-N solar observations, at the highest currently observed activity level. A medium activity case is selected for BJD 2457716 to 2457816 (2016-Nov-23 to 2017-Mar-3). A low stellar activity case is selected during the extended minimum Cycle 24 for BJD 2458950 to 2459050 (2020-Apr-10 to 2020-Jul-19). The selected chunks are shown in Fig. \ref{fig:chunks_years} with vertical dashed lines. From here onward, only the observed HARPS-N radial velocities will be considered in the analysis. As a reminder, given the matching method summarised in Section \ref{sec:match}, bad-weather breaks are already included. We note that this already significantly reduced dataset will very likely still represent an ideal stellar cadence. The Sun is still observed even with predicted SNR values down to 200. This is not the case with EPRV targets. Although the two cuts described in Section \ref{sec:HN} will eliminate data taken under not ideal conditions, these requirements are still more relaxed than what would be expected of a EPRV target. Moreover, we have not considered the possibility of telescope time competition. At the TNG, every hour of light is dedicated uniquely to the solar telescope. Conversely at night multiple programs are competing for time. It is therefore unlikely for a telescope with multiple programs to be able to achieve the "once-a-day" cadence here selected over the entire season. Nevertheless, we have reduced the dataset considerably to a cadence similar to what new missions with dedicated Earth-like targets such as the Terra Hunting Experiment \citep[THE:][]{Thompson2016} are aiming to achieve.

As mentioned in Section \ref{sec:estimate}, the uncertainties on the mean longitudinal magnetic field derived directly from SDO/HMI errors are very small. No existing or planned polarimetric survey of far-away stars could reach those levels. To better represent the stellar case, we instead inflate the uncertainties of \Bl to the best achieved uncertainty level  of 0.2 G on fully detected mean longitudinal fields for Sun-like star by the BCool collaboration \citep{Marsden2014, Mengel2017}. This is an optimistic floor that has been proven to be achievable by polarimetric observations of Sun-type stars before, and it is the precision level new spectropolarimetric instruments, such as the one currently in construction for the upcoming HARPS3, aim to achieve. We therefore use a constant error on \Bl measurements of 0.2 G. To match this uncertainty and to truly represent the inflated error, we also inject into the dataset white noise randomly extracted from a Gaussian distribution with a $\sigma$ of 0.2 G.

\subsection{Preliminary Analysis}
\label{sec:chunks}
\begin{figure}
    \centering
    \includegraphics[width=\columnwidth]{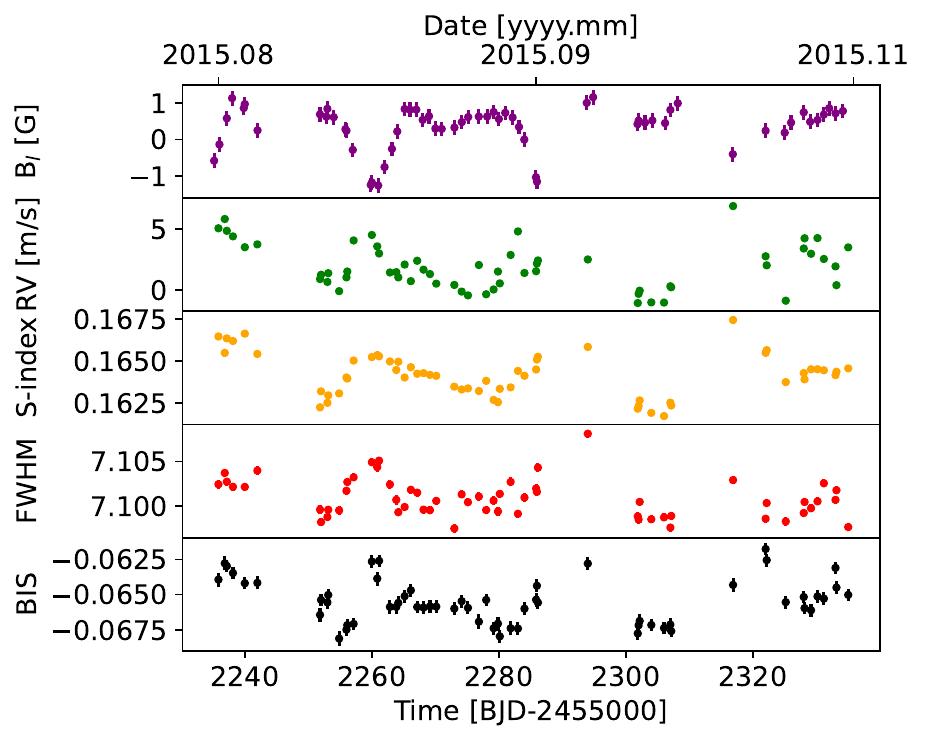}
    \caption{High activity dataset selection. From top to bottom: mean longitudinal magnetic field, HARPS-N radial velocities, S-index, FWHM, and bisector span. Some uncertainties may be too small to be clearly visible.}
    \label{fig:high_data}
    \vspace{3mm}
    \includegraphics[width=8cm]{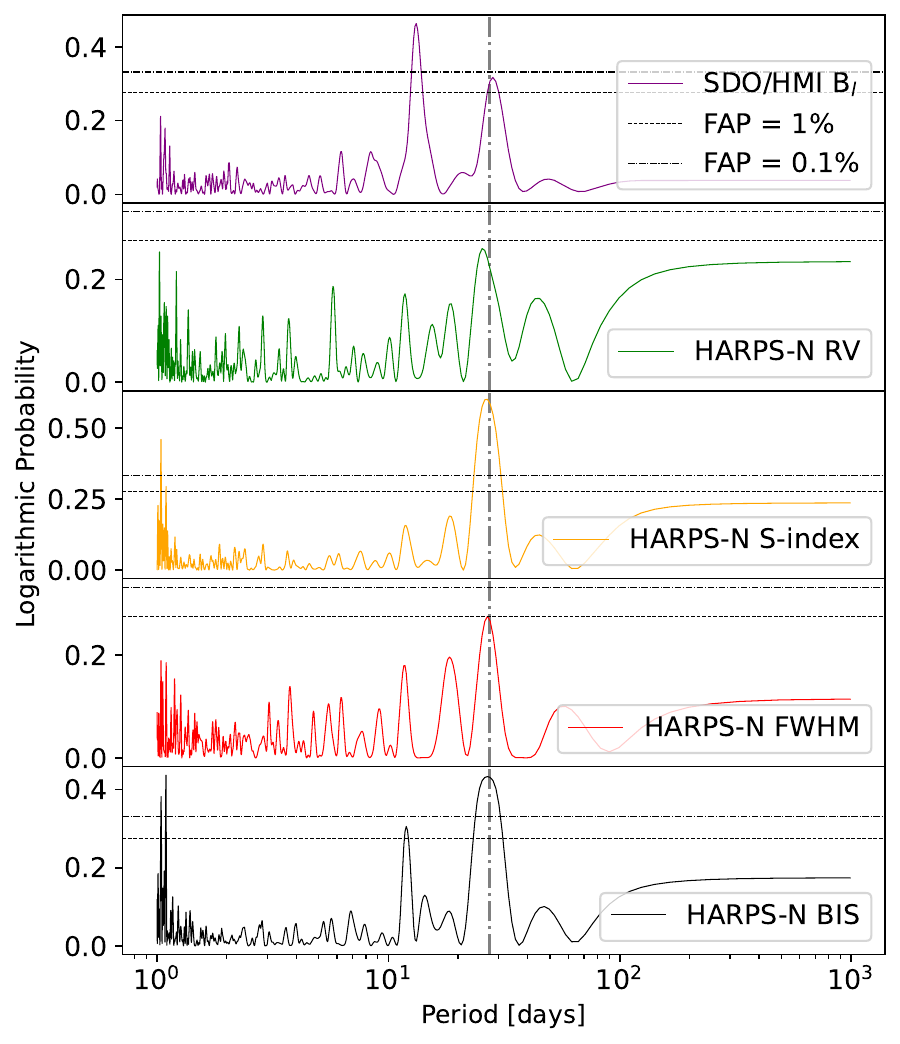}
    \caption{GLS periodograms of the high activity data. From top to bottom: mean longitudinal magnetic field, HARPS-N radial velocities, S-index, FWHM, and BIS. 1\% and 0.1\% False Alarm Probabilities are shown as dotted and dashed gray lines. The vertical dash-dotted black line highlights the Carrington solar rotational period.}
    \label{fig:high_per}
\end{figure}
\begin{figure}
    \centering
    \includegraphics[width=\columnwidth]{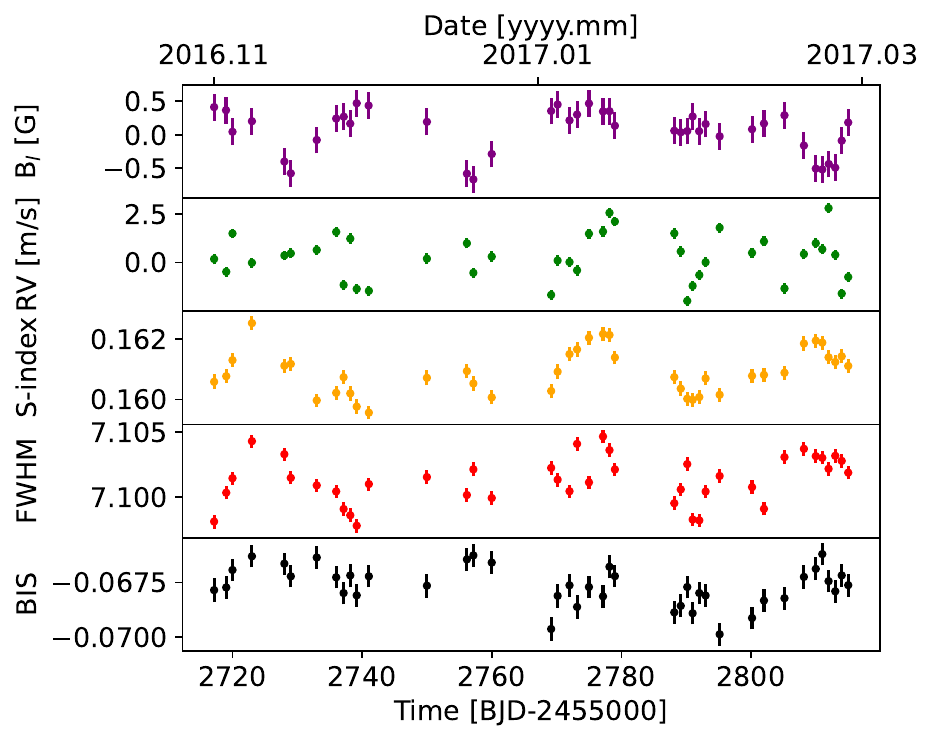}
    \caption{Medium activity dataset selection. From top to bottom: mean longitudinal magnetic field, HARPS-N radial velocities, S-index, FWHM, and BIS. Some uncertainties may be too small to be clearly visible.}
    \label{fig:med_data}
    \vspace{3mm}
    \includegraphics[width=8cm]{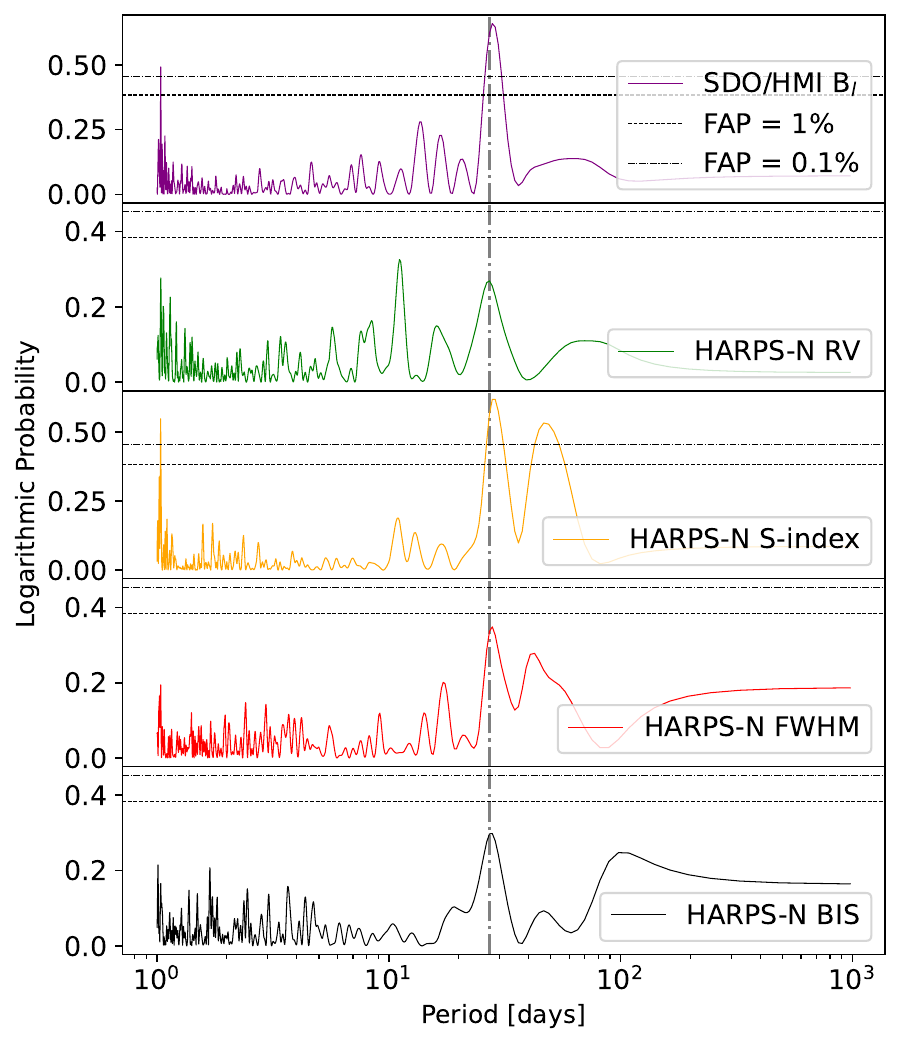}
    \caption{GLS periodograms of the medium activity data. From top to bottom: mean longitudinal magnetic field, HARPS-N radial velocities, S-index, FWHM, and BIS. 1\% and 0.1\% False Alarm Probabilities are shown as dotted and dashed gray lines. The vertical dash-dotted black line highlights the Carrington solar rotational period.}
    \label{fig:med_per}
\end{figure}
    \begin{figure}
    \centering
    \includegraphics[width=\columnwidth]{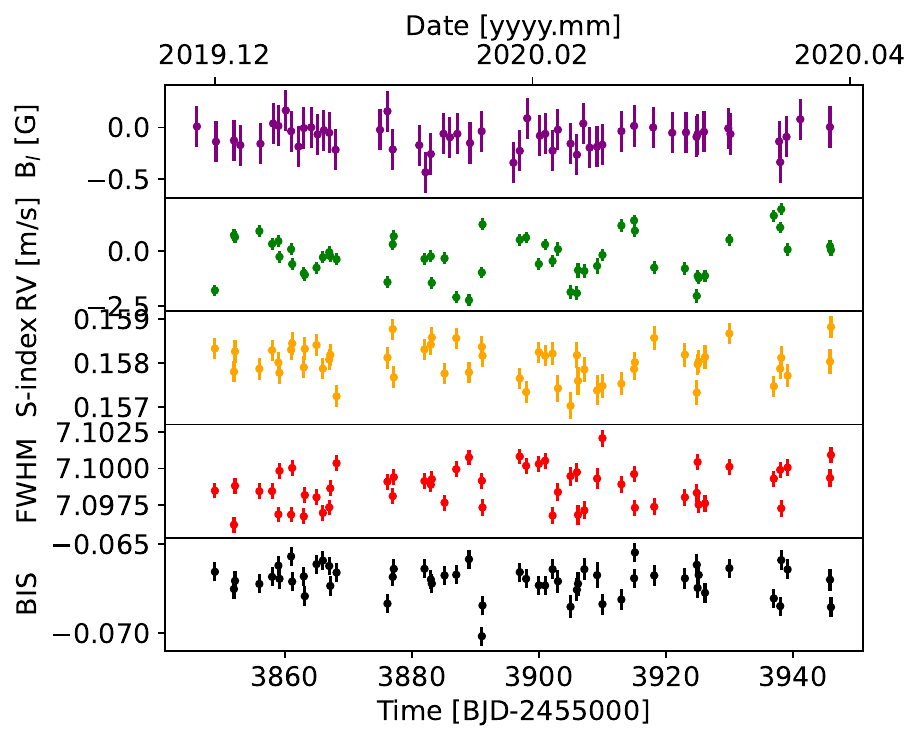}
    \caption{Low activity dataset selection. From top to bottom: mean longitudinal magnetic field, HARPS-N radial velocities, S-index, FWHM, and BIS. Some uncertainties may be too small to be clearly visible.}
    \label{fig:low_data}
    \vspace{3mm}
    \includegraphics[width=8cm]{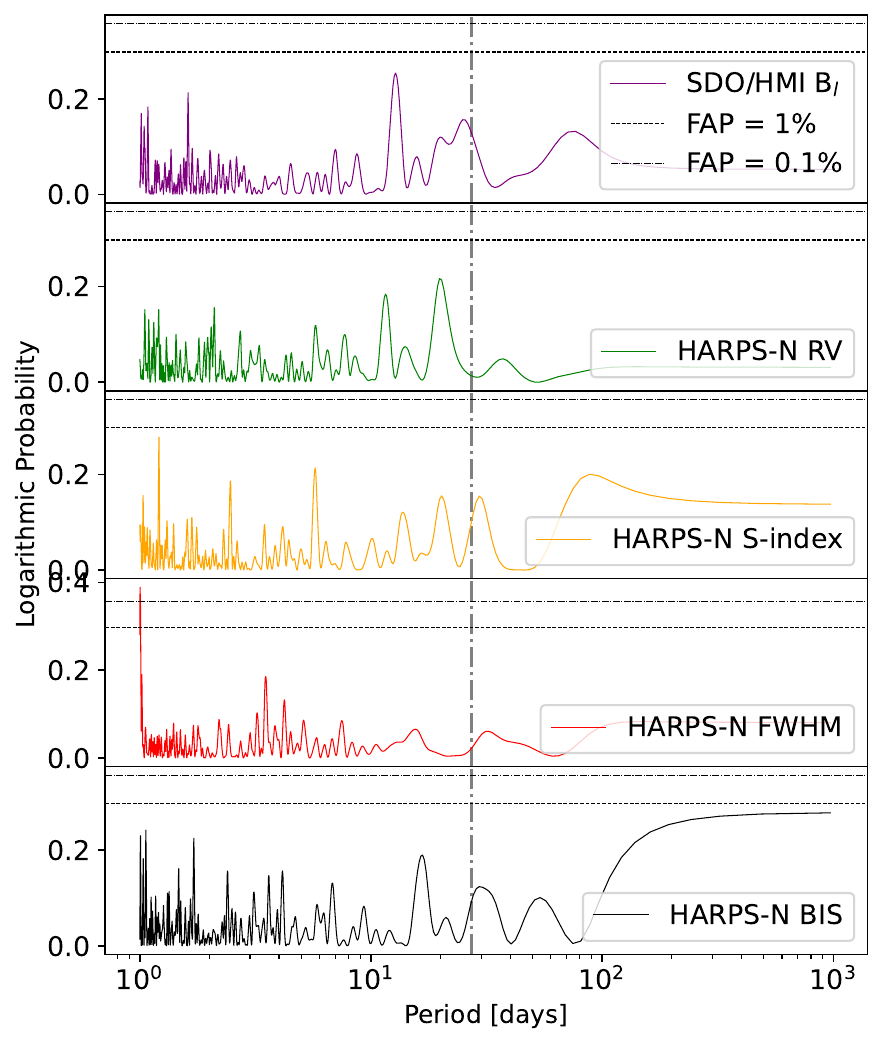}
    \caption{GLS periodograms of the low activity data. From top to bottom: mean longitudinal magnetic field, HARPS-N radial velocities, S-index, FWHM, and BIS. 1\% and 0.1\% False Alarm Probabilities are shown as dotted and dashed gray lines. The vertical dash-dotted black line highlights the Carrington solar rotational period.}
    \label{fig:low_per}
\end{figure} 
\textbf{High Stellar Activity Case:} We start with the chronologically-first 100-day chunk: the high stellar activity case. Over this window, we match 69 datapoints. The selected data is plotted in Fig. \ref{fig:high_data}. Even with a quick visual inspection, a clear periodic signal can be identified in the mean longitudinal magnetic field. Similarly to Section \ref{sec:full}, we compute the correlation between each considered activity indicator and the HARPS-N RVs, shown in the first row of Fig. \ref{fig:chunk_corr}. At this stage the solar activity is strong and dominated by rotationally-modulated effects, as shown by the high correlation between the RVs and the indicators S-index, FWHM and bisector span. We then compute the autocorrelation function of each of the considered time series, as shown in the Appendix in Fig. \ref{fig:ACF_high}. In this case, the magnetic field ACF does show a clear peak at the solar rotation period, and is the only time series for which a period can be systematically retrieved. Of the activity proxies, only the S-index includes hints to the $\sim$ 27 days period, but the peaks are too wide for a proper rotation period analysis.
We also plot the Generalised Lomb-Scargle periodograms of the all the time series for this chunk in Fig. \ref{fig:high_per}. In this case, as expected, all HARPS-N proxies and to a lower degree the radial velocities themselves have power at the solar rotation period. Once again the \Bl is sensitive to $P_{\rm rot}$ and $P_{\rm rot}/2$ signals.\\

\textbf{Medium Stellar Activity Case:}
A similar analysis is then undertaken for the medium activity case. Over this window we match 47 datapoints. We plot the derived time series in Fig. \ref{fig:med_data}. The correlation relationships between the RVs and the activity indicators are plotted in the second row of Fig. \ref{fig:chunk_corr}. The computed Spearman rank correlation coefficient are now significantly lower for all proxies and no correlation above 0.5 can now be found. We can postulate that most of the rotationally modulated effects are now reduced in significance with respect to other photospheric and chromospheric variability. As before, we compute the ACF, plotted in the Appendix in Fig. \ref{fig:ACF_med}. While some information regarding the solar rotation period could be extracted from the autocorrelation function of \Bl, at this stage of activity no systematic period extraction can be applied to any of the considered timeseries. We also perform a GLS periodogram analysis, as shown in Fig. \ref{fig:med_per}. The activity signal is now not strong enough to be picked out from a periodogram analysis of the RVs only, but it is present in most of the investigated indicators. The FWHM retrieve the rotation period to a False Alarm Probability of 1\%. The S-index and \Bl are the only ones that present power at the rotational period over the 0.1\% FAP level. It is of note that all HARPS-N activity indicators are now also presenting a peak at $\sim$40 days (not an harmonic of the rotational period or one of its aliases). This peak exceeds the 0.1\% FAP level in the S-index, the most reliable of the common proxies in the previous analysis, making period determination only based on its information trickier.
Overall at medium activity, \Bl already starts to outperform other proxies in this preliminary analysis.\\

\textbf{Low Stellar Activity Case:} We repeat the same analysis once more with the last selected dataset over the extended solar minimum. We match 79 datapoints. The extracted time series are plotted in Fig. \ref{fig:low_data}. At this level of activity the considered uncertainty of \mbox{0.2 G} is comparable to the \Bl signal itself. There is now no correlation between any of the activity proxies and the radial velocities, as shown in the bottom row of Fig. \ref{fig:chunk_corr}. Most of the rotationally-modulated effects are now overshadowed by other sources of activity.
Both an autocorrelation function (included in the Appendix in Fig. \ref{fig:ACF_low}) and a periodogram analyses yield no information regarding the periodicity of the Sun. Most of the signal seems to in fact be aperiodic, as illustrated in Fig. \ref{fig:low_per}.

\subsection{Gaussian Process Regression Analysis}
The usual next step in the analysis of radial-velocity data, especially in cases with high correlation between indicators and RVs is to employ Gaussian process regression to model the activity in the stellar proxies. This is done in order to identify the hyperparameters that better fit the stellar signal, which can then be used to inform priors in a second GP analysis of the RVs themselves. In this work, we undertake the most uninformative Gaussian process regression analysis in order to simulate a preliminary analysis, or a "worse case scenario", in a typical exoplanet detection. To model the stellar activity we use a Quasi-Period (QP) kernel \citep{Haywood2014}, with an added white noise "jitter" term in the form:
\begin{equation}\label{eq:quasiperiod}
    k(t_i,t_j) = A^2 \cdot \exp\left[ -\frac{|t_i - t_j|^2}{\tau^2} - \frac{\sin^2 \left(\frac{\pi \cdot |t_i - t_j|}{P_{\rm rot}} \right)}{\mu^2} \right] + \delta_{i,j}\beta^2,
\end{equation}
where $k(t_i,t_j)$ is the covariance function between times $t_i$ and $t_j$, $A$ is the amplitude of the signal, $\tau$ is the timescale over which the quasi-periodicity evolves (related to the evolution timescale of active regions), $P_{\rm rot}$ is the period of the stellar rotation, and $\mu$ is the harmonic complexity of the fit. $\beta$ is a jitter term and is modelling the white noise contribution to the data derived from their inherent precision. It is only applied to the diagonal of the matrix (via the Kronoker Delta function $\delta_{i,j}$).

The QP kernel has been successfully employed to model stellar activity in both radial-velocity (e.g., \citealt{Rajpaul2015, Barros2020}) and stellar activity proxy analyses (e.g., \citealt{Haywood2014, Grunblatt2015, Dalal2024}). In this work, we test whether a similar analysis could be undertaken with \Bl and whether it could be more successful than the same analysis on other activity proxies. To do so we use {\sc MAGPy\_RV}\footnote{Available at: \url{https://magpy-rv.readthedocs.io/en/latest/}} \citep{Rescigno2024, Rescigno2023code}. {\sc MAGPy\_RV} is a GP regression pipeline with affine invariant Markov Chain Monte Carlo (MCMC) parameter searching algorithm. We run the same analysis for all time series: \Bl, HARPS-N radial velocities, S-index, FWHM and bisector span. All hyperparameters are bound by forced positive (larger than 0) uniform priors. The harmonic complexity $\mu$ is bound by a uniform prior between [0,1]. We also bind both the period $P_{\rm rot}$ and the evolution timescale with uniform priors between [0,100] given the length of the selected window. We bind the white noise $\beta$ with a Gaussian prior centred in the mean value of the uncertainties of the considered dataset and of width equal to 25\% of said value, in order to avoid the GP explaining all the variability in the form of white noise. These priors represent the amount of information we are able to derive from initial analysis in the low activity case. For ease of comparison, we use the same priors in all runs. For all analyses, we simultaneously evolve 200 chains over 50,000 iterations each, with a discarded burn-in phase of 10,000 steps. We assess the convergence of the chains by computing the Gelman-Rubin statistics and define a chain as converged only under a 1.1 convergence cut. Not all chains are able to converge with the described  priors. Instead of aiming for full convergence, we select an investigation length over which most of the parameter space for all datasets is investigated, and over which all \Bl chains are fully converged. We note that no detrending has been done to remove the magnetic cycle signal from any of the datasets. While we are aware of the presence of an overall descending trend in the RVs and some of the HARPS-N activity proxies due to the decline of the magnetic activity, this effect is not especially visible in the shortened 100-days baseline. This method is meant to reproduce the results of the same uninformed process of a first analysis of stellar data. We therefore do not want to introduce any pre-whitening due to information that are not directly derived from the selected datasets.

\begin{figure}
    \centering
    \includegraphics[width=\columnwidth]{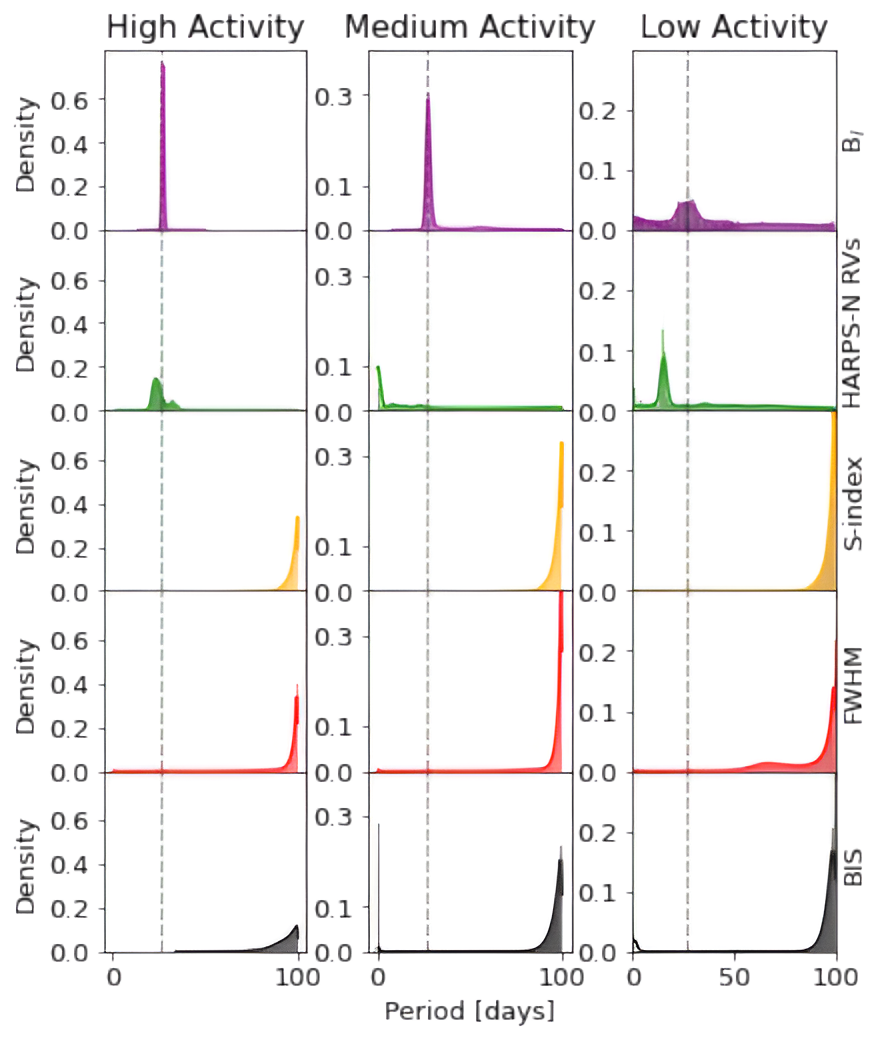}
    \caption{Collection of posteriors for the period $P_{\rm rot}$ of the Quasi-periodic kernel after GP regression. From left to right we consider the high, medium and low activity cases. From top to bottom we see the posteriors of mean longitudinal magnetic field, HARPS-N radial velocities, S-index, FWHM, and BIS in their respective colours. The Carrington solar rotation is here highlighted with a black dashed line. Note the shared y-axis for each column.}
    \label{fig:P_post}
\end{figure}
\begin{figure}
    \centering
    \includegraphics[width=\columnwidth]{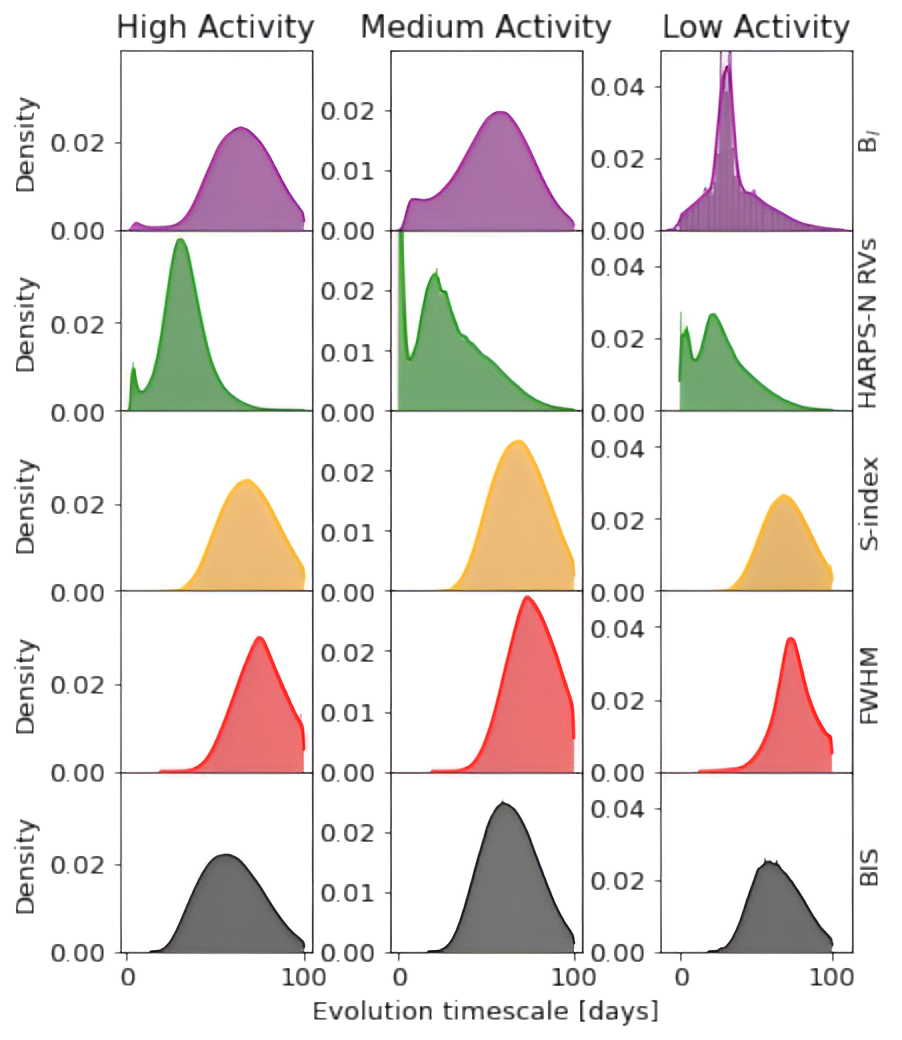}
    \caption{Collection of posteriors for the evolution timescale after $\tau$ of the Quasi-periodic kernel GP regression. From left to right we consider the high, medium and low activity cases. From top to bottom we see the posteriors of mean longitudinal magnetic field, HARPS-N radial velocities, S-index, FWHM, and BIS in their respective colours. Note the shared y-axis for each column.}
    \label{fig:evo_post}
\end{figure}
\begin{figure}
    \centering
    \includegraphics[width=\columnwidth]{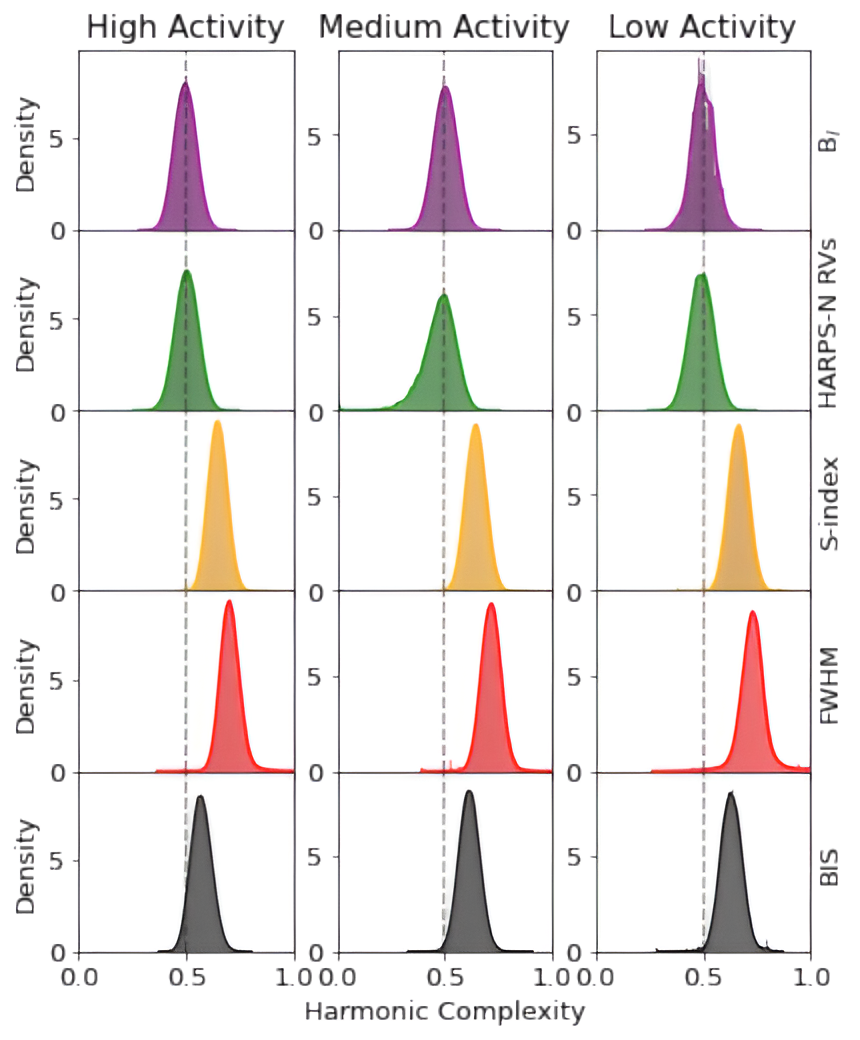}
    \caption{Collection of posteriors for the harmonic complexity $\mu$ of the Quasi-periodic kernel after GP regression. From left to right we consider the high, medium and low activity cases. From top to bottom we see the posteriors of mean longitudinal magnetic field, HARPS-N radial velocities, S-index, FWHM, and BIS in their respective colours. The black dashed line highlights the 0.5 harmonic complexity. Note the shared y-axis for each column.}
    \label{fig:harm_post}
\end{figure} 

In this work we focus on the hyperparameters useful for subsequent radial-velocity fitting: the period of the solar activity signal, its evolution timescale and its harmonic complexity. The amplitudes and jitters of each time series are not comparable. We plot the posterior distributions after MCMC analysis of each considered hyperparameter in Figs. \ref{fig:P_post}, \ref{fig:evo_post} and \ref{fig:harm_post}. The high solar activity case is shown in the first column, the medium is in the middle column, and the low activity case is in the third column. Each time series is plotted on a different row.\\

\textbf{The Rotational Period:}
We start our assessment from the period of the solar activity, in Fig. \ref{fig:P_post}. For the highest activity case the mean longitudinal magnetic field far outperforms all other proxies and the RVs themselves in identifying the "correct" solar activity period (here defined again by the Carrington solar rotational period and shown by a dashed vertical line in the figure). Therefore, even in the case in which \Bl is comparable to other common proxies in a simple Fourier analysis, the mean longitudinal magnetic field gains an edge in a GP regression framework.
At medium activity, \Bl was still able to cleanly converge for the expected value. Even during the prolonged minimum, although not to high precision, \Bl is the only time series able to identify the solar rotational period, and the radial-velocities are only sensitive to the half-period. When looking at all the posterior results together, it is clear that only \Bl is consistently successful at recovering the solar rotation period. To do so, it requires little to no prior information, making it more versatile, and can converge much quicker than any other proxy, lowering the computational expense.
In all cases, the HARPS-N activity proxies are unable to converge for any periodicity and instead their posteriors peak at the top of the time window available for exploration: 100 days. They model the activity in the data as a long period (longer than the dataset) with shorter evolution timescale $\tau$ and higher $\mu$. They therefore "assign" more of the signal to other time-dependent hyperparameter. 
This is another confirmation of the sensitivity of \Bl to the solar rotational period. Radial velocities and their spectra- or CCF-derived proxies rely on surface features and limb darkening modulation to pick up the rotational period. On the other hand, the mean longitudinal magnetic field extracted with spectropolarimetric observations is not only affected by limb-darkening and fore-shortening, but its change in intensity with rotation is also exacerbated by the fact that we are observing the line-of-sight component of the radial field, which will be at the largest when the field is pointing directly at the observer and will approach zero when rotating perpendicular to the line of sight, all together yielding a larger and clearer modulation in the signal.\\

\textbf{The Evolution Timescale:}
Similarly to the period, we plot the posterior distributions of the evolution timescale $\tau$ in Fig. \ref{fig:evo_post}. All common activity proxies as well as the mean longitudinal magnetic field prefer longer evolution timescales than the radial velocity. The $\tau$ posterior distributions of the RVs peak at values comparable to the rotation period. This result is also supported by previous analysis \citep[e.g.,][]{Camacho2022}. The longer timescale recovered by \Bl does not oppose the result of the RV analysis. In fact, even if we assume that the main source of the variations in \Bl is the magnetic flux in the active regions, the magnetic fields concentrated in said active regions have been shown to have longer lifetimes than their photometric expression in the form of sunspots or faculae (which are the source of the RV variations). A nascent active region (before any photometric brightening or dimming of the solar photosphere) is an ensemble of small-scale emergence events with a preferred magnetic orientation \citep{Strous1999}. After the emerging of magnetic field concentrations, the Ca II intensity begins increasing, usually with a time lag in the order of tens of minutes \citep{Bumba1965}, and convective collapse begins. It is then reasonable for the S-index also to converge to comparable evolution timescales to \Bl. Opposite magnetic polarities separate after 24 hours and areas of the same polarity migrate towards each other to coalesce into larger features such as pores \citep{vanDrielGesztelyi2015}. With increasing total field and as further areas migrate and conglomerate, pores evolve in photometric active regions in the form of spots and faculae \citep{Cheung2017}. With time (and with a lifetimes of 30 to 60 days) photometric active regions gradually disappear. In this process opposite polarity fragments magnetically reconnect and the flux slowly cancel itself. By the time coronal heating decreases, the plage (the coronal counterpart of faculae) start dimming. Finally the magnetic active region dissipates into the magnetic background. Just as the lifetime of spots and faculae depends on their size, the overall lifetime of magnetic active region is proportional to the magnetic flux it reaches at maximum development \citep{vanDrielGesztelyi2015}. The results of $\tau$ in \Bl can therefore be reliably employed as upper bounds in following RV GP regression analyses.\\

\textbf{The Harmonic Complexity:}
When looking at the posterior distributions for the harmonic complexity $\mu$, shown in Fig. \ref{fig:harm_post}, we note how \Bl is consistently in better agreement with the radial velocity than other proxies. The dashed vertical line indicates an harmonic complexity equal to 0.5. This $\mu$ value yields a covariance that prefers GP models with one extra "bump" per period. In a very simplistic view this could be physically equated to two active regions on opposite sides of the solar sphere with respect to the observer rotating in and out of view, or to an active region distribution that produces a similar signal. This result is in line with the conclusions in \cite{Jeffers2009}. However, we caution that the harmonic complexity is the only hyperparameter of the quasi-periodic kernel that cannot be reliably and systematically tied to a specific physical property \citep{Nicholson2022}.
From the formulation of the QP kernel in Eq. \ref{eq:quasiperiod}, a higher value of $\mu$ means a smoother curve in-between periods, or a lower inner-period complexity. These results seem to contradict the conclusions of the analysis in Section \ref{sec:full}, that the mean longitudinal magnetic field exhibits less complex signals than the other proxies. However first, the posterior distributions in Fig. \ref{fig:harm_post} need to be considered within the larger context of the Gaussian Process regression. We cannot do a direct comparison of the extracted best-fit parameters between \Bl and the HARPS-N activity proxies, as the latters were not able to recover the "correct" period and are therefore modelling the activity in a completely different manner. For example note that for all levels of activity, the best-fit jitter term extracted by the GP after MCMC optimisation is consistently larger (ranging between twice to 20 times as large) than the average uncertainty in the corresponding time series. This is not the case for \Bl. We therefore postulate that a significant part of the signal in the proxies is not being modelled by the GP at all and it is instead accounted for by the large jitter.\\

Overall, with this analysis we show that the mean longitudinal magnetic field is a great rotational period detector. It is more effective than the RVs themselves or all other considered activity proxies, as it consistently outperforms them over all solar activity levels. It is more efficient than the other considered time series, as it requires the least amount of prior information to converge the the "correct" value and needs the shortest computational time. A GP regression analysis of \Bl is not only useful to find or confirm the period of the quasi-periodic variations, but the results of other hyperparameters can also inform a second GP analysis of the RVs themselves. The harmonic complexity posterior of the mean longitudinal magnetic field can be used as a prior for the RVs, as we have proven that they are in agreement over all activity levels. Moreover, the evolution timescale derived for \Bl can inform the upper bound of the same hyperparamter for the RVs.

\section{Conclusions}
\label{sec:conclusion}
In this work we analysed the solar mean longitudinal magnetic field as a rotational period detector and as a tracer for the mitigation of activity-induced variations in RV surveys in the context of exoplanet detection. We considered the longitudinal magnetic field extracted from SDO/HMI observations alongside the $\Delta$RVs derived with a model from the same data. We performed correlation analysis, we computed  their structure functions, Generalised Lomb-Scargle periodograms and autocorrelation functions, and tested for the presence of any time lag between the two time series.
In parallel we duplicated all our analyses with Sun-as-a-star observations taken by the HARPS-N spectrograph (we considered the HARPS-N derived radial velocities, the S-index, and the full-width at half-maximum and the bisector span of the CCF). We find the following:
\begin{itemize}
    \item \Bl does not directly correlate to the RVs. This lack of correlation is not activity-level dependent. \Bl cannot therefore be employed as a direct proxy of the solar activity in the radial velocities. We however find that the RMS of \Bl computed over a window comparable to the solar rotation period do correlate well with the RVs smoothed over the same amount of time. With a rudimentary sine function fitted to the RMS of \Bl and subtracted from the SDO/HMI $\Delta$RVs, we are able to reduce the radial-velocity scatter by more than 60\%. \Bl can therefore be used to successfully model out the long-term RV signal due to the magnetic cycle of the Sun.
    \item \Bl has a significantly simpler structure function than all other considered time series, with a characteristic timescale of $\sim$10 days.
    \item  \Bl is an effective solar rotation detector. Even when the same cadence and baseline are considered between the SDO/HMI and the HARPS-N data, the periodogram of \Bl only presents peaks at the Solar Carrington rotational period and to a lesser extent to at its first and second harmonics. None of the other considered proxies or either of the radial velocities are as simple, and in most cases no clear rotation period can be isolated. This point is further confirmed by the autocorrelation analysis, in which the rotational period signal of \Bl stays clear and strong over multiple rotations. 
    \item A lag analysis was performed and a minorly relevant lag between \Bl and the RVs was found at circa $-7.5$ days. This lag appears to be driven by the signal generated by active regions. These results however, change significantly based on which section of the solar RVs are considered and based on the level of the magnetic cycle.
\end{itemize}

Overall, we therefore have shown that with high cadence and a long baseline, the mean longitudinal magnetic field is a very effective solar rotational period detector, and it can be used to inform our understanding of the physical processes happening on the surface of the Sun. This is, however, not representative of the type of observational time series we have for exoplanet detection. Therefore, we also tested the \Bl as a "stellar activity tracer". We reduced the time series to 100-days windows, with a single observation per night, and inflated the uncertainties in \Bl to those achieved by previous polarimetric surveys. We then performed a typical preliminary analysis followed by a Gaussian process regression with a quasi-periodic kernel. We performed the same analysis for three chunks of data over high, medium and low activity levels. We found the following:
\begin{itemize}
    \item The mean longitudinal magnetic field outperforms the other activity indicators in a preliminary periodogram analysis for the medium and low activity levels.
    \item After one-dimensional GP regression using a quasi-periodic kernel, \Bl is the only time series (compared against HARPS-N RVs, S-index, FWHM and BIS) that is able to successfully recover the "correct" rotational period over all levels of activity. It does so with the shortest convergence time and with little to no prior information.
    \item \Bl (as well as all other proxies) recovers a longer evolution timescale than the RVs.
    \item The best-fit harmonic complexities of \Bl and the RVs strongly agree within uncertainties.
\end{itemize}

With this analysis, we have reconfirmed the mean longitudinal magnetic field as an effective and efficient rotational period detector, with exoplanet-survey-like time series and over all levels of solar activity. The best-fit values extracted from the posteriors of the other hyperparameters can be used as prior information for a follow-up RV GP regression. This work also highlights the need of time series of polarimetric data for less magnetically active stars, in more fields than simply exoplanetology.

\section*{Acknowledgements}
The author would like to thank the referee for the feedback provided in the review stage.

FR is funded by the University of Exeter's College of Engineering, Maths and Physical Sciences, UK.

The HARPS-N project was funded by the Prodex Program of the Swiss Space Office (SSO), the Harvard University Origin of Life Initiative (HUOLI), the Scottish Universities Physics Alliance (SUPA), the University of Geneva, the Smithsonian Astrophysical Observatory (SAO), the Italian National Astrophysical Institute (INAF), University of St. Andrews, Queen’s University Belfast, and University of Edinburgh.

This work has been carried out within the framework of the NCCR PlanetS supported by the Swiss National Science Foundation under grants 51NF40\_182901 and 51NF40\_205606.
XD acknowledges the support from the European Research Council (ERC) under the European Union’s Horizon 2020 research and innovation programme (grant agreement SCORE No 851555) and from the Swiss National Science Foundation under the grant SPECTRE (No 200021\_215200).

RDH is funded by the UK Science and Technology Facilities Council (STFC)'s Ernest Rutherford Fellowship (grant number ST/V004735/1).

SD acknowledges support from the STFC consolidated grant number ST/V000721/1.

BSL is funded by a UK Science and Technology Facilities Council (STFC) studentship (ST/V506679/1).

XD acknowledges funding by the French National Research Agency in the framework of the Investissements d'Avenir program (ANR-15-IDEX-02), through the funding of the "Origin of Life" project of the Grenoble-Alpes University.

ACC acknowledges support from STFC consolidated grant numbers ST/R000824/1 and ST/V000861/1.

BK acknowledges funding from the European Research Council under the European Union’s Horizon 2020 research and innovation programme (grant agreement No 865624, GPRV)

S.H.S. gratefully acknowledges support from NASA XRP grant 80NSSC21K0607 and NASA EPRV grant 80NSSC21K1037.

NP acknowledges the Scholarship program funded by the Knut and Alice Wallenberg foundation.

\section*{Data Availability}
The observational data included in this publication are all either currently or soon-to-be public. The SDO/HMI images are available at \url{https://sdo.gsfc.nasa.gov/data/aiahmi/}. The first half of the HARPS-N solar data is available at \url{https://dace.unige.ch/observationSearch/?observationType=[%22solarSpectroscopy%22]}. The second half is in the process of release and will be described and made available in an upcoming publication.



\bibliographystyle{mnras}
\bibliography{mybib} 




\appendix

\section{Correlation Plots}
\begin{figure*}
    \centering
    \includegraphics[width=15cm]{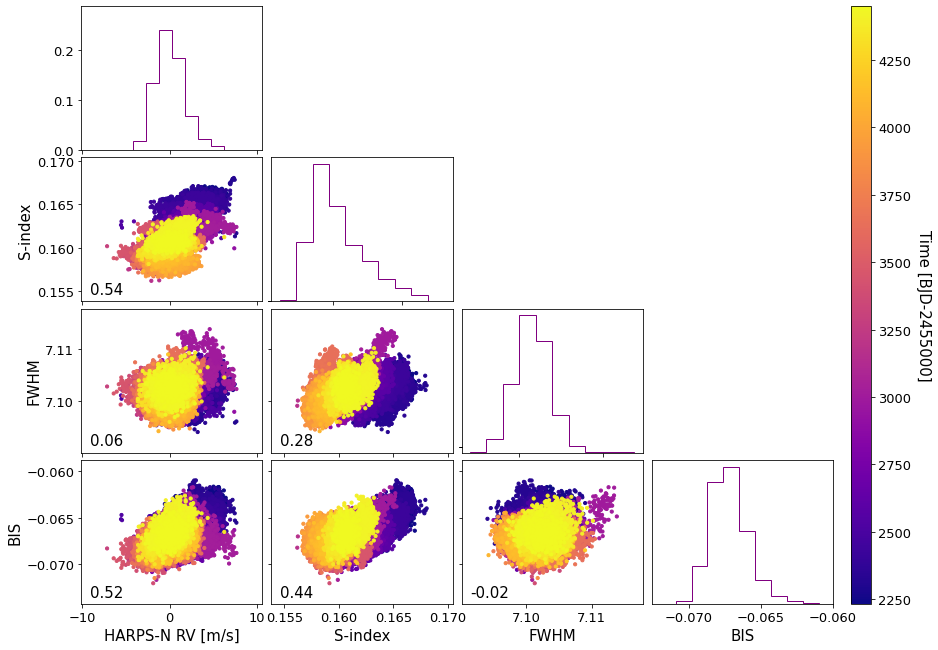}
    \caption{Correlation plot between the HARPS-N radial velocities and their activity proxies, S-index, full width at half maximum, and bisector span. The color bar indicates the BJD of each datapoint. The Spearman Rank correlation factor for each set is also included.}
    \label{fig:HN_corr}
\end{figure*}
We include in this appendix the correlation plots between the HARPS-N Sun-as-a-star solar radial velocities and the activity proxies considered in the paper: the spectra-derived S-index, and the CCF-derived full width at half maximum and bisector span. The correlations are plotted in Fig. \ref{fig:HN_corr}. The data is colour-coded with increasing BJD. In each panel we also include the computed Spearman correlation coefficient rank. More information can be found in Section \ref{sec:full_corr}.

\section{A Justification for the  Dataset Matching Method} \label{sec:appendix_matching}

There are in principle a number of approaches to the problem of combining time series with unequal sampling.
The approach we take is to linearly interpolate the SDO/HMI data onto the timestamp of the closest HARPS-N observation, provided that these two observations are separated by no more than one hour.
If there are no HARPS-N observations within one hour of a given SDO/HMI observation, that data point is omitted.
In contrast to interpolating the SDO/HMI data onto every HARPS-N timestamp, each SDO/HMI observation is being considered at most once.
This avoids spurious correlations in Section \ref{sec:full} where a single SDO/HMI data point corresponds to many HARPS-N observations.\\

\begin{figure}
    \centering
    \includegraphics[width=\linewidth]{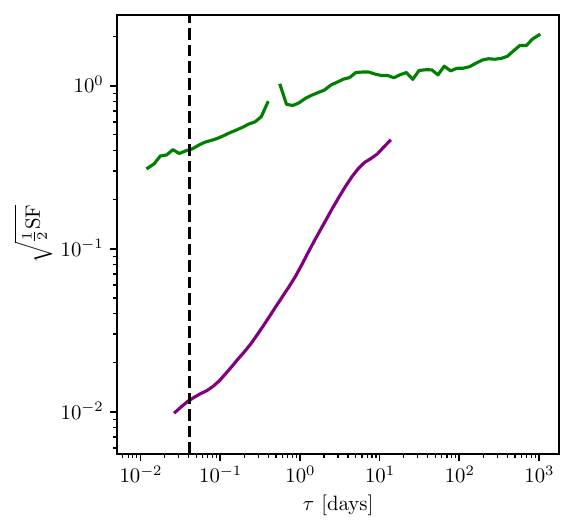}
    \caption{Structure functions of the HARPS-N RVs (green) and mean longitudinal magnetic field (purple). Note that the longitudinal magnetic field used here is a higher cadence time series than the one used in the main text. A vertical dashed line indicates one hour, the maximum interpolation distance in the matching procedure used here. The greater steepness of the purple line indicates that interpolating the SDO/HMI data introduces less spurious signal than interpolating the HARPS-N data.}
    \label{fig:high_cadence_sf}
\end{figure}

It may appear counter intuitive to interpolate the sparser data set onto the data set with the denser sampling. 
To justify this approach, we produced a high-cadence time series from every available set of 720s-exposure SDO/HMI images from 1-Jan-2017 to 1-Feb-2017.
This month was chosen as it represents roughly the median activity level we explore.
In Fig. \ref{fig:high_cadence_sf}, we plot the structure functions of both the HARPS-N RVs and the high-cadence mean longitudinal magnetic field\footnote{See the main text for more details}.
The steeper slope of the structure function of the mean longitudinal magnetic field demonstrates that there is significantly less fractional variability within the longitudinal magnetic field time series at timescales of less than an hour than there is in the HARPS-N RVs.
We are therefore justified in interpolating the sparser time series onto the denser grid, as opposed to the inverse.
To further assess the strength of any possibly injected signals via interpolation, we also compute the mean absolute deviation (MAD) of the difference between all consecutive SDO/HMI observations. The derived MADs are 0.074 \ms for the SDO/HMI $\Delta$RVs and 0.015 G for \Bl. These values represent the average dispersion of the separation between each subsequent observations and are also small enough to be negligible for the scope of this work.

\section{Autocorrelation Function Analysis of the Stellar-like Observations}
\begin{figure}
    \centering
    \includegraphics[width=\linewidth]{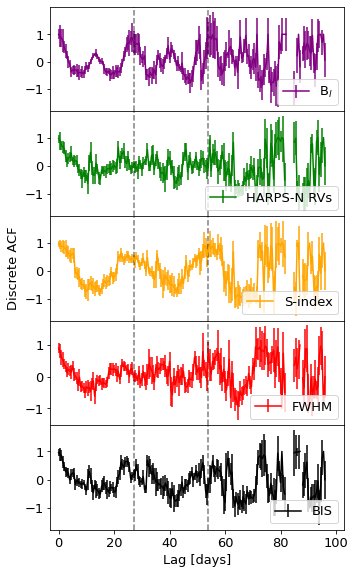}
    \caption{Autocorrelation function of the matched time series of the high-activity section of from top to bottom: longitudinal magnetic field in purple, the HARPS-N RVs in green, the HARPS-N S-index in orange, the HARPS-N FWHM in red and the HARPS-N BIS in black. Uncertainties are included in the form of errorbars. The vertical dashed lines in gray represent (from left to right) the approximated solar rotation period ($\sim$27 days) and twice that value ($\sim$54 days).}
    \label{fig:ACF_high}
\end{figure}

\begin{figure}
    \centering
    \includegraphics[width=\linewidth]{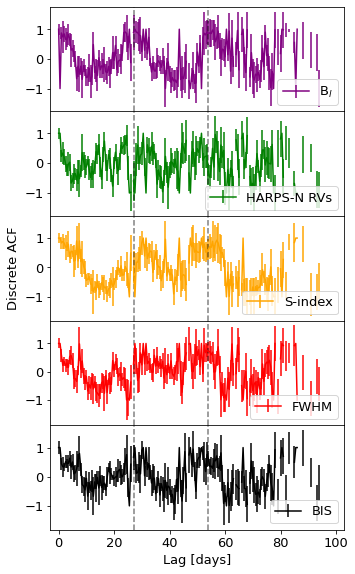}
    \caption{Autocorrelation function of the matched time series of the medium-activity section of from top to bottom: longitudinal magnetic field in purple, the HARPS-N RVs in green, the HARPS-N S-index in orange, the HARPS-N FWHM in red and the HARPS-N BIS in black. Uncertainties are included in the form of errorbars. The vertical dashed lines in gray represent (from left to right) the approximated solar rotation period ($\sim$27 days) and twice that value ($\sim$54 days).}
    \label{fig:ACF_med}
\end{figure}

\begin{figure}
    \centering
    \includegraphics[width=\linewidth]{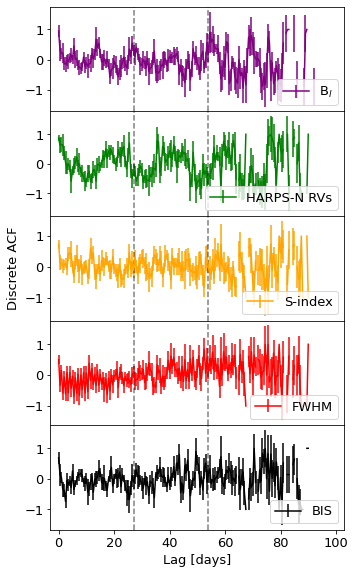}
    \caption{Autocorrelation function of the matched time series of the low-activity section of from top to bottom: longitudinal magnetic field in purple, the HARPS-N RVs in green, the HARPS-N S-index in orange, the HARPS-N FWHM in red and the HARPS-N BIS in black. Uncertainties are included in the form of errorbars. The vertical dashed lines in gray represent (from left to right) the approximated solar rotation period ($\sim$27 days) and twice that value ($\sim$54 days).}
    \label{fig:ACF_low}
\end{figure}

We include in this appendix the autocorrelation function plots of all the considered time series in Section \ref{sec:sub} for the three selected levels of activity: the matched longitudinal magnetic field, HARPS-N RVs, S-index, FWHM and BIS. All the ACFs are plotted in Figs. \ref{fig:ACF_high}, \ref{fig:ACF_med} and \ref{fig:ACF_low} for respectively the high, medium and low activity data sections. In each panel we also include the solar rotation period and its second pulse at respectively $\sim$27 and $\sim$54 days (the Carrington period and twice that value) as dray vertical dashed lines. The only ACF that can be used to reliably inform us about the solar rotation period is the ACF of the high-activity time series of \Bl, but is all other cases the peaks due to the rotation are either too wide or not present at all.

\bsp	
\label{lastpage}
\end{document}